\documentclass[useAMS,usenatbib]{mn2e}
\usepackage{graphicx} 
\usepackage{float}
\usepackage{times}
\usepackage{epsfig}
\usepackage{epstopdf}
\usepackage{rotating}
\usepackage{sidecap}
\usepackage{longtable}
\usepackage{lscape}
\usepackage{color}
\usepackage{wrapfig}

\def\lesssim{\mathrel{\hbox{\rlap{\hbox{\lower4pt\hbox{$\sim$}}}\hbox{$<$}}}}
\let\la=\lesssim
\def\gtrsim{\mathrel{\hbox{\rlap{\hbox{\lower4pt\hbox{$\sim$}}}\hbox{$>$}}}}
\let\ga=\gtrsim

\title[QPOs and states in NS X-ray binaries]{Links between quasi-periodic oscillations and accretion states in neutron star low mass X-ray binaries}
\author[S.E. Motta et al.]{S.E. Motta$^{1, 2}$, A. Rouco-Escorial$^{2, 3}$, E. Kuulkers$^{2,4}$, T. Mu\~noz-Darias$^{5,6}$, A. Sanna$^{7}$ \\
$^{1}$University of Oxford, Department of Physics, Astrophysics, Denys Wilkinson Building, Keble Road, OX1 3RH, Oxford, United Kingdom\\
$^{2}$ESAC, European Space Astronomy Centre, Villanueva de la Ca\~nada, E-28692 Madrid, Spain\\
$^{3}$Anton Pannekoek Institute for Astronomy, University of Amsterdam, Postbus 94249, 1090 GE Amsterdam, The Netherlands\\
$^{4}$ESTEC, European Space Research and Technology Centre, Keplerlaan 1, NL-2200 AG Noordwijk, the Netherlands\\
$^{5}$ Instituto de Astrof\'isica de Canarias, 38205 La Laguna, Tenerife, Spain \\
$^{6}$ Departamento de astrof\'isica, Univ. de La Laguna, E-38206 La Laguna, Tenerife, Spain \\
$^{7}$Dipartimento di Fisica, Universit\`a degli Studi di Cagliari, SP Monserrato-Sestu km 0.7, 09042 Monserrato, Italy}

\begin{document}
\maketitle
\begin{abstract}

\noindent We analysed the Rossi X-ray Timing Explorer (RXTE) data from a sample of bright accreting neutron star (NS) low-mass X-ray binaries (LMXBs). With the aim of studying the quasi-periodic variability as a function of the accretion regime, we carried out a systematic search of the quasi-periodic oscillations (QPOs) in the X-ray time series of these systems, using the integrated fractional variability as a tracker for the accretion states.
We found that the three QPO types originally identified in the '80s for the brightest LMXBs, the so-called Z-sources, i.e., horizontal, normal and flaring branch oscillations (HBOs, NBOs and FBOs, respectively), are also identified in the slightly less bright NS LMXBs, the so-called Atoll sources, where we see QPOs with a behaviour consistent with the HBOs and FBOs.
We compared the quasi-periodic variability properties of our NS sample with those of a sample of black hole (BH) LMXBs. We confirm the association between HBOs, NBOs and FBOs observed in Z-sources, with the type-C, type-B and type-A QPOs, respectively, observed in BH systems, and we extended the comparison to the HBO-like and FBO-like QPOs seen in Atoll sources. We conclude that the variability properties of BH and weakly-magnetized NS LMXBs show strong similarities, with QPOs only weakly sensitive to the nature of the central compact object in both classes of systems.
We find that the  historical association between kHz QPOs and high-frequency QPOs, seen around NSs and BHs, respectively, is not obvious when comparing similar accretion states in the two kinds of systems. 

\end{abstract}

\begin{keywords}
Neutron star - accretion disc - binaries: close - X-rays: stars
\end{keywords} 
\section{Introduction}\label{Sec:intro}

Low-mass X-ray binaries (LMXBs) are accreting systems harbouring a compact object, either a neutron star (NS) or a black hole (BH), that is fed mass by a Sun-like star companion. Accretion takes place through an accretion disk, where matter is heated through viscosity to temperatures high enough (about 10$^7$K) to radiate in the X-ray band (\citealt{Shakura1973}). 
Based on their long-term behaviour, LMXBs can be classified into persistent and transient sources. While persistent sources are active all the time, showing X-ray luminosities typically above L$_X$ $\sim$ 10$^{36}$ erg/s, transient sources spend most of their life in a dim, quiescent state (L$_X$ $\sim$ 10$^{30}$--10$^{34}$ erg/s), occasionally interrupted by outbursts lasting $\sim$weeks to months, with recurrence times of months to decades. Apart from a few exceptions, BH systems are normally transients, while NS systems account for the vast majority of the persistent sources population. 

Detailed and extended studies of BH LMXBs in outburst performed over the last decades have established that they all follow a similar outburst evolution, displaying a number of spectral-timing states clearly connected with different accretion regimes onto the black hole (e.g. \citealt{Miyamoto1992}, \citealt{Belloni2016}). During the outburst rise and at the end of their  active phase, transient BH systems typically show an \textit{hard} state, during which the energy spectrum is dominated by Comptonized emission and the fast-time variability level is high (more than 20-30 percent of the total emission is variable). The hard state precedes a fast transition (lasting $\sim$days) to a \textit{soft} state, that takes the source across an \textit{intermediate} state at a roughly constant luminosity (about tens percent the Eddington Luminosity L$_{Edd}$). 
During the soft state the energy spectrum is dominated by the thermal emission from the accretion disk and the fast-time variability is very low. During the major state transition from hard to soft state, very distinctive changes in the fast-time variability occur, contrasting sharply with rather smooth changes in the energy spectra, which evolve from Compton emission-dominated to disk emission-dominated in a rather continuous fashion.  
At this point of the outburst evolution, shortly before or just after the transition to the soft state, a few sources enter the so called ultra-luminous state (ULS, e.g. \citealt{Motta2012}). GRO J1655-40 is the BH LMXB that showed the clearest example of ULS in the RXTE era. Other sources have shown short excursions to this state, among these GX 339-4, XTE J1550-564, H1743-322, 4U 1630-47 (see e.g. \citealt{Dunn2010}) and GRS 1915+105, which is almost all the time observed in such a state (\citealt{Munoz-Darias2014}). The ULS  shows fairly low variability, but large color variations (spectral transitions), and, compared to the standard intermediate states, it extends to much higher luminosities, reaching in some cases the Eddington Luminosity (\citealt{Uttley2015}).
After the soft state is reached, the luminosity slowly decreases (in weeks to months) to a few percent of L$_{Edd}$ until a backward transition occurs (also in this case at fairly constant flux), taking the source through the intermediate state and then to the hard state, after which the active phase normally comes to an end. 
Outburst phases in BH LMXBs are characterized by hysteresis, i.e. the hard to soft transition always occurs at higher luminosity than the soft to hard transition (\citealt{Miyamoto1995}). This cyclic pattern translates into hysteresis loops in the hardness-intensity diagram (HID, \citealt{Homan2001}) and rms-Intensity diagram (RID, \citealt{Munoz-Darias2011a}). 

The phenomenology associated to NS LMXBs is somewhat more complex than that found in BH systems, likely because of the presence of a hard surface and an anchored magnetic field, both bound to influence the accretion of matter in the vicinity  of the NS. NS LMXBs show different spectral states that are in many ways similar to those seen in BH systems (see \citealt{VDK2006} for a review). Based on the shape of their Color-Color diagram (CCD, \citealt{Hasinger1989}), NS LMXBs are classified as \textit{Atoll} and \textit{Z} sources. Atoll sources show three main states - the \textit{island}, the \textit{lower banana} and the \textit{upper banana} - roughly corresponding to the hard, intermediate and soft state, respectively, seen in BH systems. Z-sources, instead, are almost always seen in a fairly soft state, and their CCDs show a typical Z-shaped track formed by three branches, dubbed as \textit{horizontal}, \textit{normal} and \textit{flaring}. 
Only a few sources, the most famous of which is the NS LMXB XTE J1701-462 (e.g. \citealt{Lin2009}), have so far shown the full transition between Z and Atoll phase, allowing to make a clear connection between Atoll and Z-sources. Other examples are EXO 1745-248 in Terzan 5 (\citealt{Barret2012}), and Cir X-1 (\citealt{Oosterbroek1995} and \citealt{Shirey1998}).

Historically, hysteresis has been associated to BH transients, even if a few NS transient systems have shown clear examples of hysteresis patterns during their outbursts (e.g. Aql X-1, \citealt{Maccarone2003}). Recently, \cite{Munoz-Darias2014} showed  that the hysteresis patterns between Compton-dominated and thermal-dominated states are also common in NS LMXBs, and the RID provides a common framework to describe accretion states in both BH and NS LMXBs. In particular, the spectral, timing and multi-wavelength properties of a given source - NS, either Atoll or Z-source, or BH LMXB - can be determined by its location in this diagram. 
Below $\sim$50\% L$_{Edd}$ (Atoll sources and most BH LMXBs) hysteresis is the natural form that state transitions take and does not seem to require large changes in luminosity to occur. Above $\sim$50\% L$_{Edd}$ (Z sources and highly accreting BHs) hysteresis does not take place: independently on the nature of the compact object (BH or NS, either bright atoll or Z-source), the systems show fast colour (\textit{horizontal}) variations (reflecting short transitions from and to the soft, thermal dominated state) and flaring behaviour, while showing fairly low total fast-time variability \citep{Munoz-Darias2014}. During the evolution across different states, the main difference that remains between BH and NS sources resides in the transition velocity between hard and soft state: while the fastest transitions are seen in Z-sources and the slowest in BH systems, Atoll sources show transition time scales somewhat in between the two. Along such transitions, the variations of the accretion rate can change significantly from source to source (see e.g. \citealt{Church2014}, \citealt{Seifina2015}).

Both when hysteresis happens and when it does not, NS and BH systems cross three main accretion states: hard state, above 20\% rms; intermediate state, between 5 and 20\% rms; soft state, below 5\% rms. In other words, while the accretion rate/luminosity determines the transition mode (hysteresis or horizontal transitions), the spectral state is directly and univocally related to the variability level (i.e., high rms always means hard state, low rms always means soft state).

\subsection{Quasi-periodic oscillations}

Quasi-periodic oscillations (QPOs) are common feature in both BH and NS LMXBs. In a Fourier power density spectrum (PDS), they take the form of relatively narrow
peaks and appear together with different kinds of broad-band noise components
(e.g. \citealt{VDK2000}). 
In both NS and BH systems QPOs have been observed in a wide range of frequencies and have been divided in low-frequency QPOs (LFQPOs, seen between few mHz up to $\sim$30\,Hz in BHs and $\sim$60\,Hz in NSs) and high-frequency QPOs (HFQPOs,  observed between 100 and 500\,Hz in BHs, and from several hundreds Hz up to $>$ 1\, kHz in NSs, where they are normally referred to as kHz QPOs). Despite the fact that both LFQPOs and HFQPOs have been known for several decades, their origin is still debated, even though several models have been put forward to explain them. LFQPOs have been interpreted in terms of relativistic effects (e.g. \citealt{Stella1999} and Ingram{2009}), as the consequence of instabilities of different kinds  (see e.g. \citealt{Lamb2001} and \citealt{Tagger1999}), or as the results of oscillations of a boundary  transition layer formed by matter  adjusting to the sub-Keplerian boundary conditions near the central compact object (see  \citealt{Titarchuk1999}). In particular, the relativistic Lense-Thirring precession seems to be, to date, the most promising mechanism for the existence of certain LFQPOs both in NS and BH LMXBs (see  \citealt{Stella1998} and \citealt{Ingram2009}).
HFQPOs are mainly explained in terms of relativistic effects linked to the orbital frequencies (\citealt{Stella1998}), or as a result of resonances between  frequencies arising in the accretion flow around a compact object (see, e.g., \citealt{Abramowicz2001} and \citealt{Abramowicz2004}, \citealt{Kato2004} and \citealt{Kato2005}).
 
LFQPOs in BH LMXBs have been classified into three types: A, B, C (e.g. \citealt{Wijnands2001}, \citealt{Casella2005}, \citealt{Motta2015}). In Z-sources, LFQPOs are classified differently depending on the branch the source occupies in the CCD when the QPO is observed. Three main types of LFQPOs have been identified:  horizontal branch oscillations (HBOs), normal branch oscillations (NBOs) and flaring branch oscillations (FBOs, \citealt{VdK1989}).
\cite{Casella2005}  compared the QPOs observed in a small sample of BH LMXBs and Z-sources, and proposed an association with FBOs, NBOs and HBOs with type-A, -B, -C QPOs, respectively. In Atoll sources, the situation is less clear. QPOs have been divided in groups depending on their position on a CCD (see e.g.\citealt{DiSalvo2003}), but a less sharp distinction of states in Atoll sources (opposed to the fairly obvious horizontal-flaring-normal branch pattern in Z-sources)  makes a classification less obvious (see e.g. \citealt{vanStraaten2001}).  \cite{Wijnands1999a} attempted to define a common classification of Atoll and Z-sources QPOs performing a systematic analysis of the data available at the time. However, the significantly smaller amount of existing data and a substantially poorer phenomenological picture available back in 1999 did not allow to define a clear classification scheme. 

HFQPOs have been classified in upper and lower kHz QPOs (see \citealt{vanderKlis1997} for a review) in NS LMXBs and, by analogy, in upper and lower HFQPOs in BH LMXBs (see e.g. \citealt{Strohmayer2001}). While in the case of BH systems HFQPOs tend to appear at high accretion luminosities and soft accretions states (\citealt{Belloni2012}), in the case of NSs the relation between the appearance of kHz QPOs and the different spectral states appears to be rather complex (see e.g. \citealt{Belloni2007} and \citealt{Zhang2017}).

The phenomenology associated to quasi periodic variability is known to be richer in NS systems than in BH systems, and QPOs that do not have an equivalent in BH LMXBs sometimes appear in PDS from NS systems. Some examples of QPOs that do not find a correspondent in BH systems are: (i) QPOs at $\sim$1\,Hz sometimes seen in dipping sources (see e.g. \citealt{Jonker1999}); (ii) hector-Hz QPOs  seen in Atoll sources (see e.g. \citealt{Altamirano2008});  (iii)  QPOs with frequencies in the mHz range, likely resulting from marginally stable nuclear burning of Hydrogen/Helium on the NS surface (e.g., \citealt{Heger2007}). 
While case (iii) is clearly related to processes occurring at the NS surface, QPOs described in (i) and (ii) are linked to the accretion flow and likely have a geometrical origin. 

Motivated by the results obtained by \cite{Munoz-Darias2014}, in this paper we analyse the quasi-periodic variability properties of a sample of NS systems spanning a large accretion rate range.
Our goal is to investigate the evolution of LFQPOs and kHz QPOs in relation to the spectral state using the total integrated fractional rms (i.e. a measure of the fast-time variability level) as an accretion regime tracker (\citealt{Munoz-Darias2011a}). We also aim at establishing a \textit{global} classification of LFQPOs in NSs to be compared to the A, B, C classification valid for BH XRBs, in an attempt to confirm (or reject) the association between LFQPOs in NSs and BHs.

\begin{figure}
\centering
\includegraphics[width=0.49\textwidth]{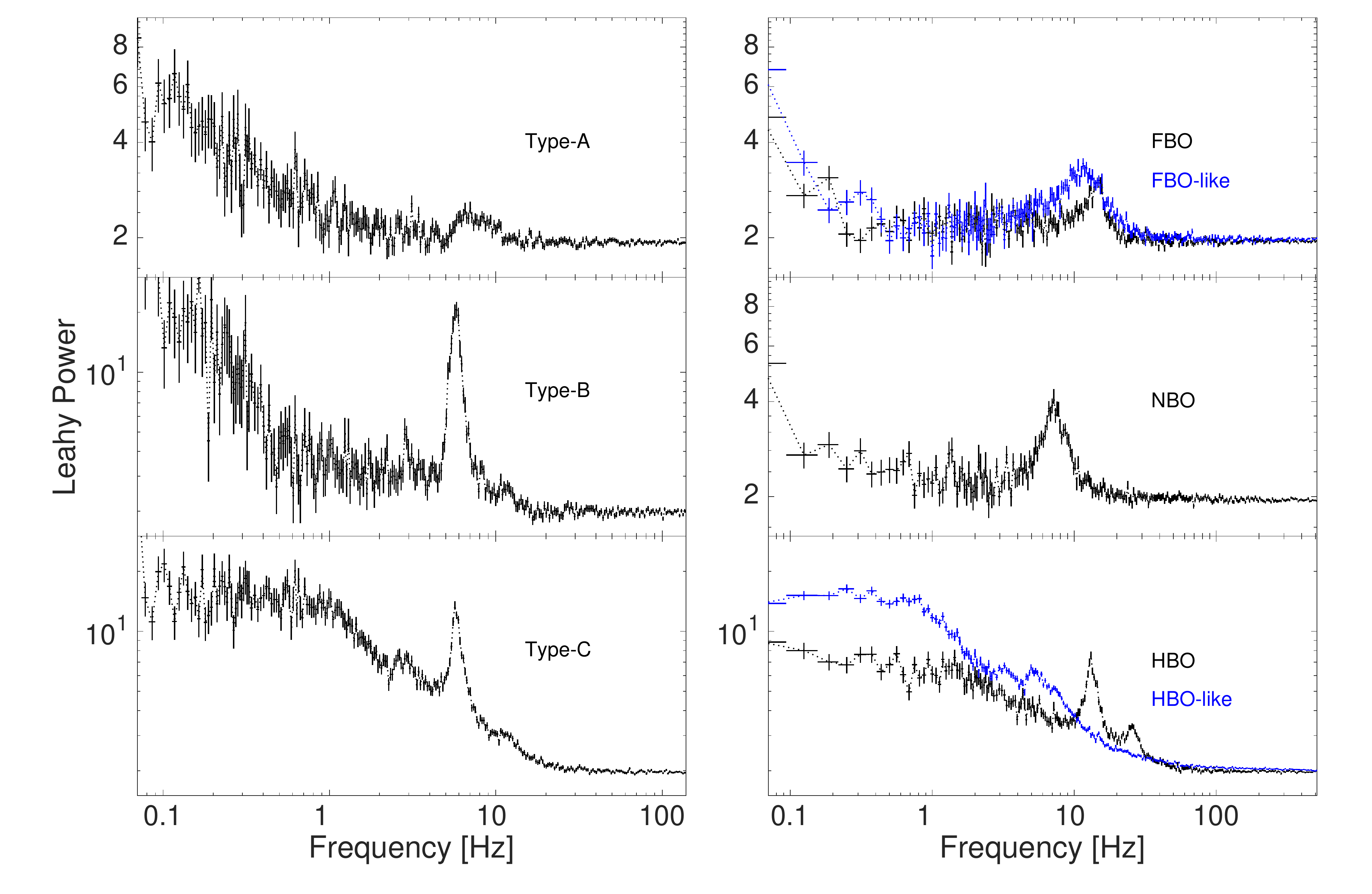}
\caption{\textit{Left  panels}: examples of QPOs from BH LMXBs. From top to bottom, QPOs are taken from  XTE J1859+226, GX 339-4 and again GX 339-4. \textit{Right panels}: examples of QPOs from NS LMXBs. From top to bottom, QPOs are taken from GX 17+2 and 4U 1705-44 (FBO and FBO-like QPOs, respectively),  GX 17+2, and Cyg X-2 and 4U 1728-34 (HBO and HBO-like QPOs, respectively). See text for the definition of HBO/FBO-like QPOs.}\label{fig:QPO_sel}
\end{figure}
\begin{table*}
\caption{List of NS sources considered in this work. We report in the table the source class (Atoll or Z-source or hybrid and subcategories, based on \citealt{Munoz-Darias2014}), the type (persistent or transient), and the type of QPOs detected, including the frequency range and rms range where they appear.}\label{tab:sources}
\begin{tabular}{c c c c c c c c}	
\hline
Source name		&			Class			&	type			& 		HBOs				& 	NBOs				&		FBOs							&	kHz QPOs			    \\
\hline
\hline	
Cyg X-2			&	Cyg-like Z				&	persistent  &   $\sim$10-56\, Hz  		& $\sim$5-6\,Hz			&	  $-$						    &	$\sim$650-750\, Hz	    \\
				&							&				&   $\sim$2-16\, rms\% 		& $\sim$2-6\, rms\% 	&	  $-$						    &	$\sim$8-12\, rms\%    \\

GX 17+2			&	Sco-like Z				&   persistent	&   $\sim$20-60\, Hz		& $\sim$6-9\, Hz	&   $\sim$10-15\,Hz					&	$\sim$600-800\,Hz	   	\\
				&							&   		    &   $\sim$2-9\, rms\% 		& $\sim$2-6 rms\% 	&   $\sim$3-6\, rms\%					&	$\sim$4-9\, rms\%   	\\

XTE J1701-462	&	hybrid (Atoll and Z)	&   persistent  &   $\sim$9-63\,Hz			& $\sim$7\,Hz		&	  $-$	   						&	$\sim$500-950\,Hz   	\\
				&							&   			&   $\sim$2-16\, rms\% 		& $\sim$2\,rms\%    &	  $-$	   						&	$\sim$3-9\, rms\%   	\\

4U 1705-44 		&	Atoll with bright Atoll & 	persistent  &		$-$					&	  $-$			&   $\sim$12-20\,Hz					&	$\sim$750-880\,Hz   	\\
		 		&						 	& 				&		$-$					&	  $-$	    	&   $\sim$5-9\,rms\%    			&	$\sim$5-7 rms\%   	\\

Aql X-1			&	Atoll					&	transient	&   $\sim$2-8\, Hz&	  $-$   &	  $-$	   		& 	$\sim$580-880\, Hz	   	\\
				&							&				&    $\sim$20\, rms\%& $-$	&	  $-$	   		& 	$\sim$2-10\,rms\%   	\\

IGR 17473-2721	&	Atoll					&	transient	& $\sim$2-4\,Hz				&	  $-$			&	  $-$	   						& 	$\sim$770-870\,Hz		\\		
				&							&				& $\sim$19-20\, rms\%    	&	  $-$			&	  $-$	   						& 	$\sim$ 3\, rms\%   		\\		

4U 1728-34		&	Atoll					&   persistent	&    $\sim$10-56\, Hz		&	  $-$			&	  $-$							&	$\sim$400-1100\, Hz		\\
				&							&   			&    $\sim$ 5-21\, rms\%   	&	  $-$			&	  $-$							&	$\sim$4-22\, rms\%   	\\

\hline
\end{tabular}

\end{table*}

\section{Observations and data analysis}\label{sec:observations} 

In this work we will only focus on the most commonly detected types of LFQPOs and kHz QPOs, which have been already associated to some extent to QPOs in BH LMXBs.
As noted above, while an established classification for LF QPOs in Z-sources exists, this is not true for Atoll sources. It is known, however, that QPOs in Atoll sources closely resemble those seen in Z-sources (e.g. \citealt{Wijnands1999a}), with the main difference that they appear to have, in general, a lower intrinsic amplitude (Fig. \ref{fig:QPO_sel}). In particular, the QPO type most often detected in Atolls show variable frequency and can therefore be associated to the HBOs in Z-sources. Furthermore, some Atolls (e.g. 4U 1705-44 and XTE J1806-246) also show QPOs that have been classified as FBOs (\citealt{Wijnands1999b}), while no QPO that could be associated to an NBO has been detected so far. Based on these facts, we classify QPOs from Atoll sources into HBO-like QPOs and FBO-like QPOs.

\subsection{Sample selection}

Our primary objective is to investigate the behaviour of QPOs in NS LMXBs in different accretion states, while performing a systematic classification of QPOs independent from the Atoll or Z nature of the system. 
\cite{Munoz-Darias2014} showed that the NS LMXBs observed by RXTE showed remarkably consistent behaviours in the RID depending on their average accretion rate (high accretion rate corresponds to horizontal fast transitions, while low accretion rates correspond to hysteresis loops, see Sec. \ref{Sec:intro}). For this reason, we can assume that all sources of the same type (i.e. Atoll sources, bright Atolls, Sco-like and Cyg-like Z sources) behave in the same way from the accretion/states point of view. We thus selected a sample of sources by choosing one or more representative system for each source type and spanning, as a whole, the largest possible luminosity/accretion rate range. In order of increasing average luminosity/accretion rate, this selection translated in a sample including the following systems:

\begin{itemize}

\item 3 Atoll sources, of which two transients (Aql X-1 and IGR 17473-2721) and one persistent (4U 1728-34). Aql X-1 and IGR 17473-2721, despite displaying only a small number of LFQPOs, show hysteresis cycles and, more in general HIDs, remarkably similar to those from BHBs (\citealt{Munoz-Darias2014}). 4U 1728-34 has been extensively monitored by RXTE, showing loops in the HID on the 30-40 days time-scale. Compared to other Atoll sources, 4U 1728-34 shows a larger number of LFQPOs (M. van Doesburgh, private communication).

\item an Atoll source with a \textit{bright Atoll} phase, 4U 1705-44 (persistent).   This source is one of the NS systems that show the clearest hysteresis loops  (\citealt{Munoz-Darias2014}), but also shows an obvious bright-atoll phase, that  connects Atoll sources to Z-sources.

\item an hybrid source, XTE J1701-462, transient. XTE J1701-462 is the first source that unmistakably showed spectral and timing properties of both Z and atoll sources (\citealt{Homan2007}).  

\item two Z-sources, Cyg X-2 and GX 17+2, both persistent systems. GX 17+2 is a Sco-like source\footnote{Sco X-1 represents a data analysis challenge due to its extreme fluxes (Sco X-1 is among the closest X-ray binaries known) that resulted in a number of complex RXTE modes, difficult to analyse consistently with all the other sources of our sample. }, accreting at rates on average slightly lower with respect to Cyg-like sources. Cyg X-2 is, by definition, a Cyg-like source, i.e. a Z source reaching the highest luminosities seen for neutron star systems.

\end{itemize}

\noindent We summarize in Tab. \ref{tab:sources} information about the NS sources of our sample and about the QPOs we detected in each of them.

\bigskip

In order to compare NS and BH systems in relation with their accretion state/quasi-periodic variability link, we selected a number of BH LMXBs based on \cite{Motta2015}. As for NS systems, also BHs show remarkably similar behaviours in the RID \citep{Munoz-Darias2014}, denoting their consistent behaviour in outburst and across their spectral-timing states. 
While BH LMXBs show accretion rates lower than NS systems, a few sources occasionally approach the Eddington limit. Therefore,  we selected sources showing different accretion rates and spanning, as a whole, the largest possible accretion rate interval. 
Additionally, since type-A QPOs and HFQPOs are fairly rare in BH systems, we also required that our sample contained all the BH LMXBs that did show type-A QPOs and HFQPOs in the RXTE era. The resulting sample includes: GX 339-4, XTE J1550-564, XTE J1859+226, XTEJ1817-33, 4U 1543-47 and GRO J1655-40. It is worth noticing that while GRO J1655-40 does not show type-A QPOs, it shows in its PDS a peaked noise component, or \textit{bump}, which is thought to be related to type-B QPOs (\citealt{Motta2012}) and is  typical of the ULS, thus worth consideration because seen also in NS systems \citep{Barret2002}.

Despite the fact that the selected BH sample is smaller than that used by \cite{Motta2015}, based on the finding of these authors we know that BH systems behave consistently also from the quasi-periodic variability point of view, showing specific QPOs spanning similar frequency ranges and appearing in the same areas of the RID (i.e. in the same spectra/timing states). Therefore, we can reasonably assume that this sample is representative for the entire class of BH LMXBs showing QPOs.

\subsection{Data analysis}\label{sec:data analysis} 

We examined all the RXTE archival data of the NS sources in our sample, for a total of 4083 observations. For each observation we computed power spectra from RXTE/PCA data using custom software under \textsc{IDL}\footnote{GHATS, http://www.brera.inaf.it/utenti/belloni/GHATS\_Package/Home.html} in the energy band 2-115 keV (absolute PCA channel 0 to 249). We used 16s-long intervals and a Nyquist frequency of 2048 Hz. We averaged the PDS and subtracted the contribution of the Poissonian noise (see \citealt{Zhang1995}). The PDS were normalized according to \cite{Leahy1983}.  We measured the total integrated fractional rms (for both NS and BH systems) integrating in the frequency range  0.1 - 64 Hz the PDS produced in the energy range 2-15 keV (RXTE/PCA absolute channels 0-35) energy band.
Count rates for the RID were also collected in the 2-15 keV energy band. 

For our subsequent analysis we selected all those observations where at least one narrow (quality factor\footnote{$Q = \nu_{centroid}/FWHM$,  where $\nu_{centroid}$ is the centroid frequency of the QPO fitted with a Lorentzian and FWHM its full width at half maximum)} Q $>$ 2) feature or a \textit{bump} could be detected on top of the broad-band noise in the PDS. 
PDS fitting was carried out within IDL through the  {\sc MPFIT} package. We fitted the broad-band noise with a variable number of broad Lorentzian model components, while LFQPOs were well-described by a variable number of narrow Lorentzian shapes depending on the presence of harmonic peaks \citep{Belloni2002}. When two or more peaks were present  (in harmonic relation to each other), we identified the fundamental based on the QPO evolution along the outburst. The harmonic properties of QPOs from  4U 1728-34 differ slightly from those of the other sources in the sample, therefore we treated these QPOs in a different way (see Appendix A for details on the analysis).  

Based on the results of the fitting, we excluded from the analysis non-significant features (significance\footnote{The significance of QPOs are given as the ratio of the integral of the power of the Lorentzian used to fit the QPO divided by the negative 1-sigma error on the integral of the power.} $\leq$ 3$\sigma$). Our final sample includes a total of 601 observations with an average exposure of 1-2 ks, whose PDS contain one or more QPO and/or a bump, for a total of 710 detections.
We fitted logarithmically binned PDS, which is ideal for the analysis of LFQPOs, but not always appropriate when hunting for HFQPOs, especially when they are marginally significant. Since our aim is to investigate the relation between QPOs and accretion states, this will not affect our results. However, we caution the reader that this study is not to be intended as a complete and exhaustive search for HFQPOs in a sample of NS systems.

Finally, we classified all the QPOs according to the HBO/HBO-like, NBO, FBO/FBO-like scheme described at the beginning of this section. 

\section{Results}

We computed the rms versus frequency diagram and the rms versus Intensity diagram (RID) for our NS sample and BH sample (see Fig. \ref{fig:rms_plots} and Fig. \ref{fig:RID}). The LFQPOs from the BH sample  shown in Figs. \ref{fig:rms_plots} (right panel), and \ref{fig:RID} (right panel) and \ref{fig:RID1655} are taken from \cite{Motta2012} and \cite{Motta2015}, while HFQPOs (Fig. \ref{fig:rms_plots_high}, right panel) are  taken from \cite{Belloni2012}, where we considered all the HFQPOs reported with a single trial significance larger than 3$\sigma$. 

\subsection{Low frequency QPOs}\label{Sec:LFQPOs}

We will initially focus on the properties and behaviour of LFQPOs in relation with the accretion states of their host sources. We will then move to kHz QPOs and HFQPOs in Sec. \ref{Sec:HFQPOs}.

\subsubsection{Rms-frequency diagrams}\label{sec:rms_vs_fr}

\begin{figure*}
\centering
\begin{tabular}{c c}
RMS VS FREQ DIAGRAM - NEUTRON STARS	& RMS VS FREQ DIAGRAM - BLACK HOLES \\
LFQPOs	& LFQPOs \\

\includegraphics[width=0.49\textwidth]{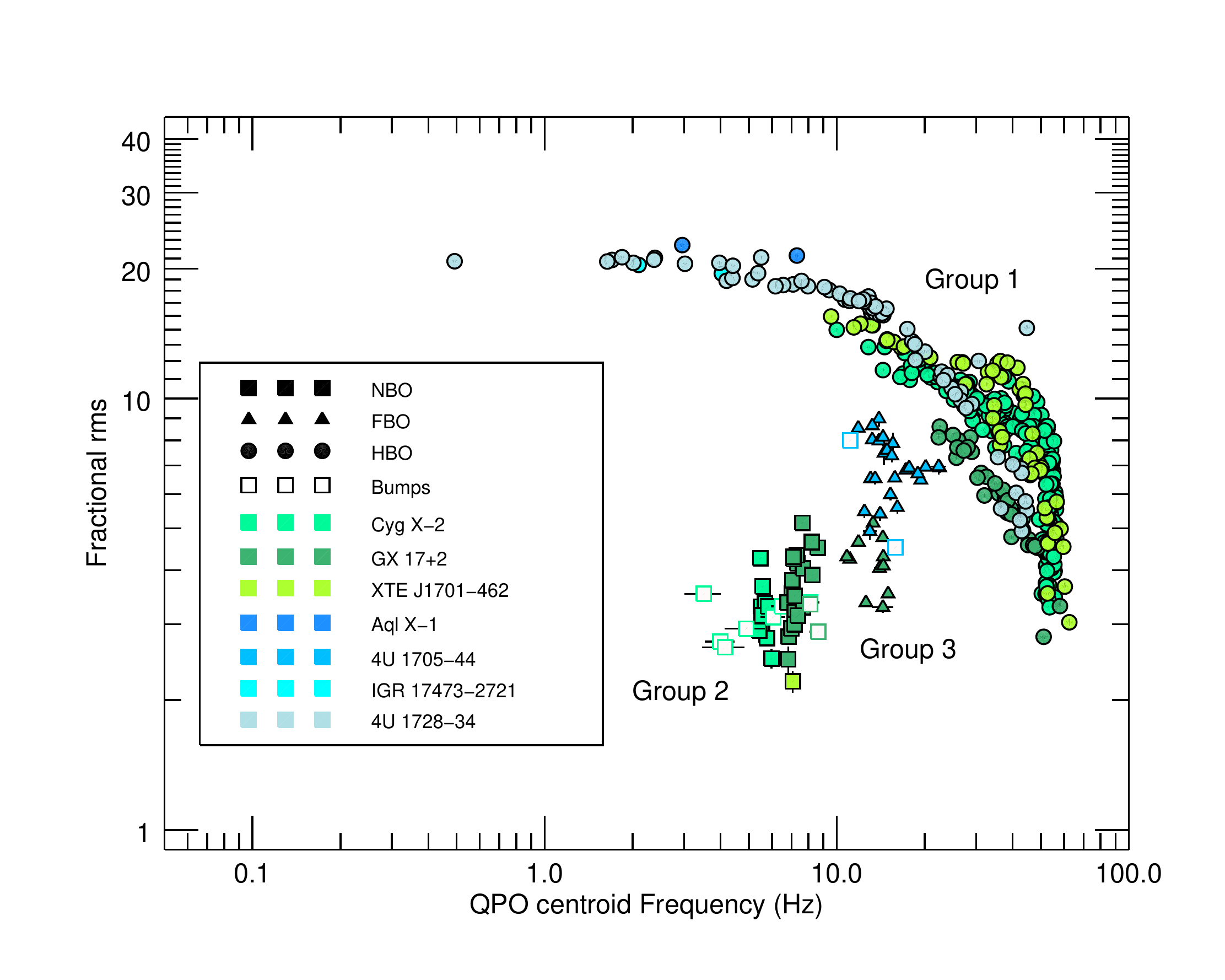} & 
\includegraphics[width=0.49\textwidth]{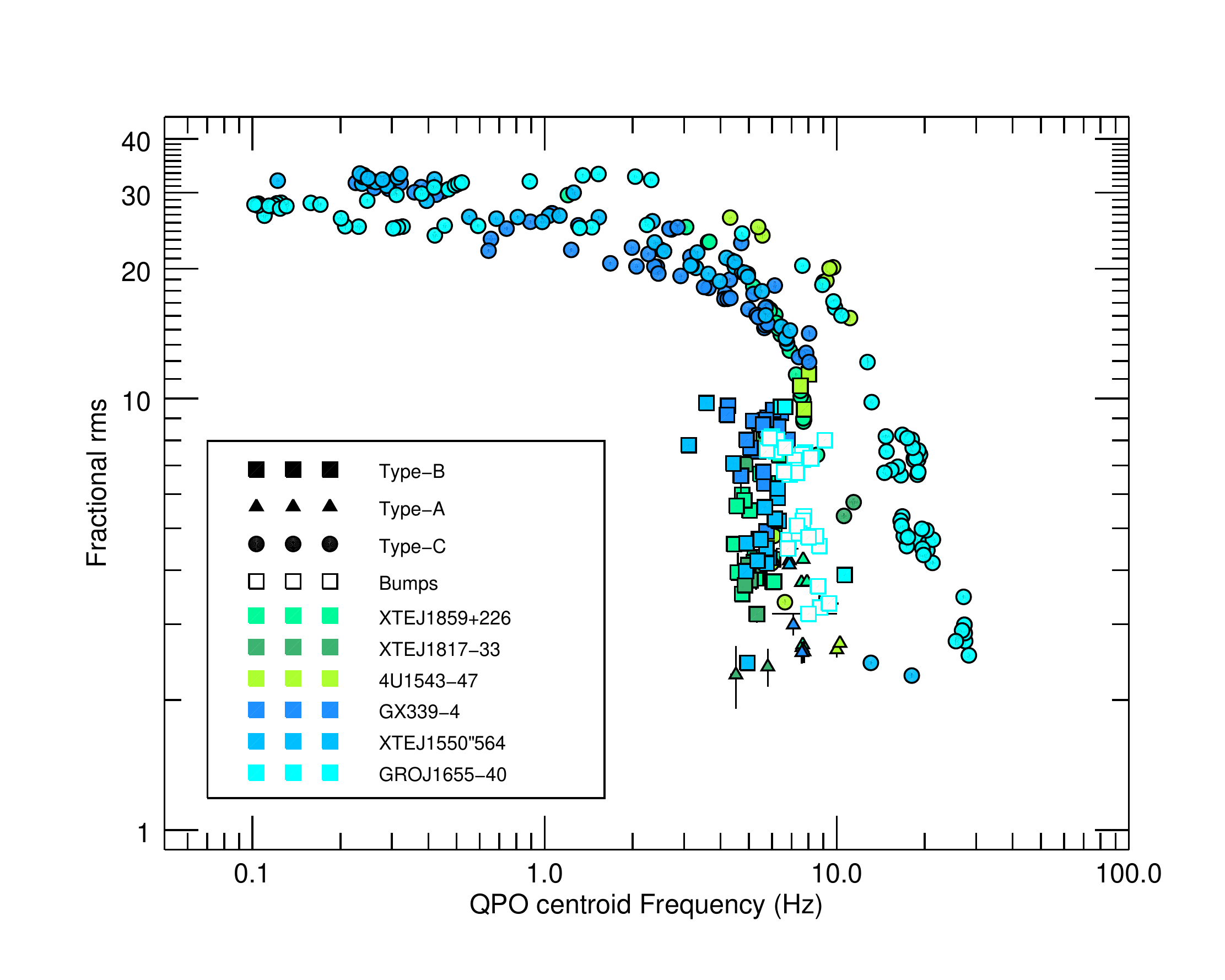} \\
\end{tabular}
\caption{\textbf{Left panel} - QPO centroid frequency versus integrated fractional rms (0.1--64.0 Hz, 2-16\, keV) from our NS sample. Each point corresponds to a single QPO while different colours  represent different sources (see legend). \textbf{Right panel} QPO centroid frequency versus integrated fractional rms (0.1--64.0 Hz, 2-16\, keV) for our BH sample. Also in this case, each point corresponds to a QPO and each colour correspond to a different source (see legend). }\label{fig:rms_plots}
\end{figure*}

Plotting the total integrated fractional rms versus the centroid
frequency of a QPO from the same observation is a useful method for distinguishing different types of QPOs, which in a rms versus frequency plot are known to generally form different, well-defined groups (see, e.g., \citealt{Casella2005} and \citealt{Motta2012}). 
Therefore, we plot the integrated fractional rms versus the centroid frequency of the QPO for each observation in our NS sample and for each QPO detected in each observation (see  Fig. \ref{fig:rms_plots}). Inspecting the dynamical PDS of observations showing FBOs it is clear that this type of oscillation can vary significantly in frequency during one observation on time scales as short as few seconds, but usually this happens around a preferred value ($\sim$13-15 Hz). This results in a peak that is broader in the average PDS (produced averaging several 16s long PDS) than in a PDS extracted in a 16s-long data snapshot. However, the peak in the average PDS is centred at the frequency where most of the QPO power is produced, which is the most relevant information for our analysis.

%
%
%
%
%
%
\begin{figure*}
\centering
\begin{tabular}{c c}
RID - NEUTRON STARS	& RID - BLACK HOLES \\
\includegraphics[width=0.48\textwidth]{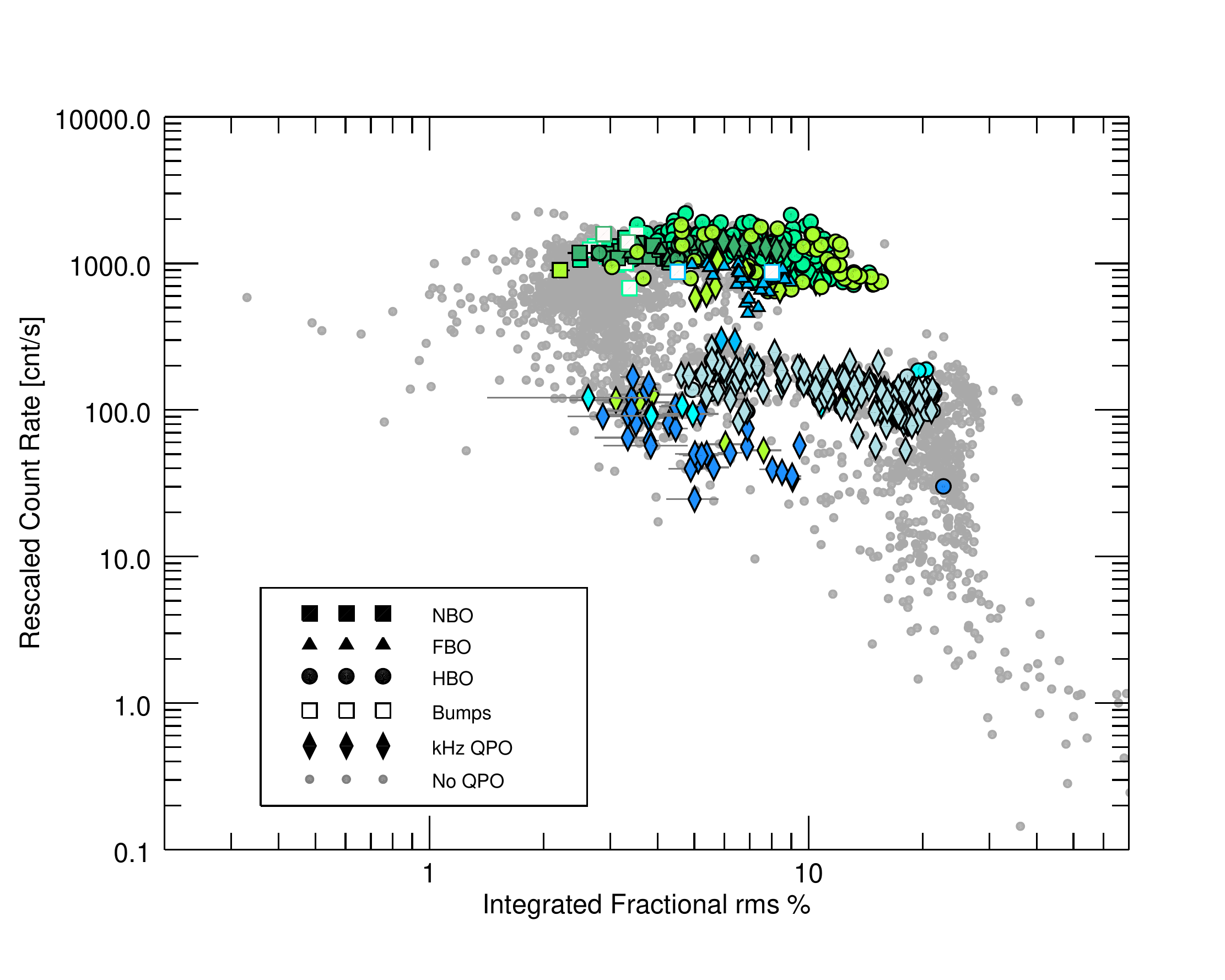} &
\includegraphics[width=0.48\textwidth]{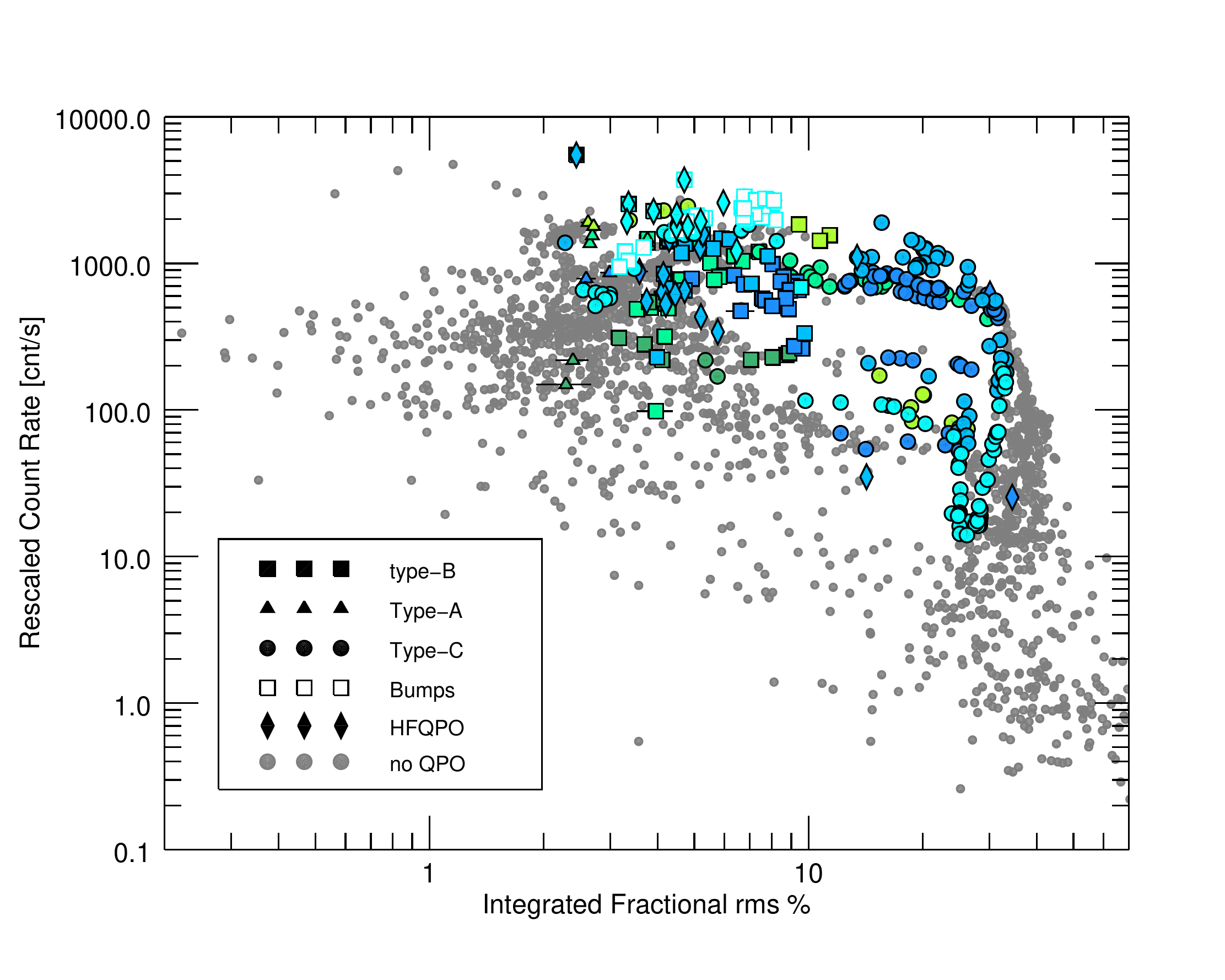} \\
\end{tabular}
\caption{\textbf{Left panel:} Rms-Intensity diagram including the observations from the 7 NS XRBs of our sample. Gray points correspond to observations where no QPO is observed, while the different symbols mark different QPO types (see Legend). The colour coding is the same used in Fig. \ref{fig:rms_plots} (left panel). All count rates were rescaled to that of Cyg X-2, see text for details. \textbf{Right panel:} rms-Intensity diagram  (RID) for BH sample. The different symbols mark the QPO types detected (see Legend). The colour coding is the same used in Fig. \ref{fig:rms_plots} (right panel). As in the case of NSs, we rescaled all the count rates at the distance of 13 kpc, see text for details.}\label{fig:RID}
\end{figure*}
\begin{figure}
\centering
\begin{tabular}{c}
RID - GRO J1655-40 \\
\end{tabular}
\includegraphics[width=0.49\textwidth]{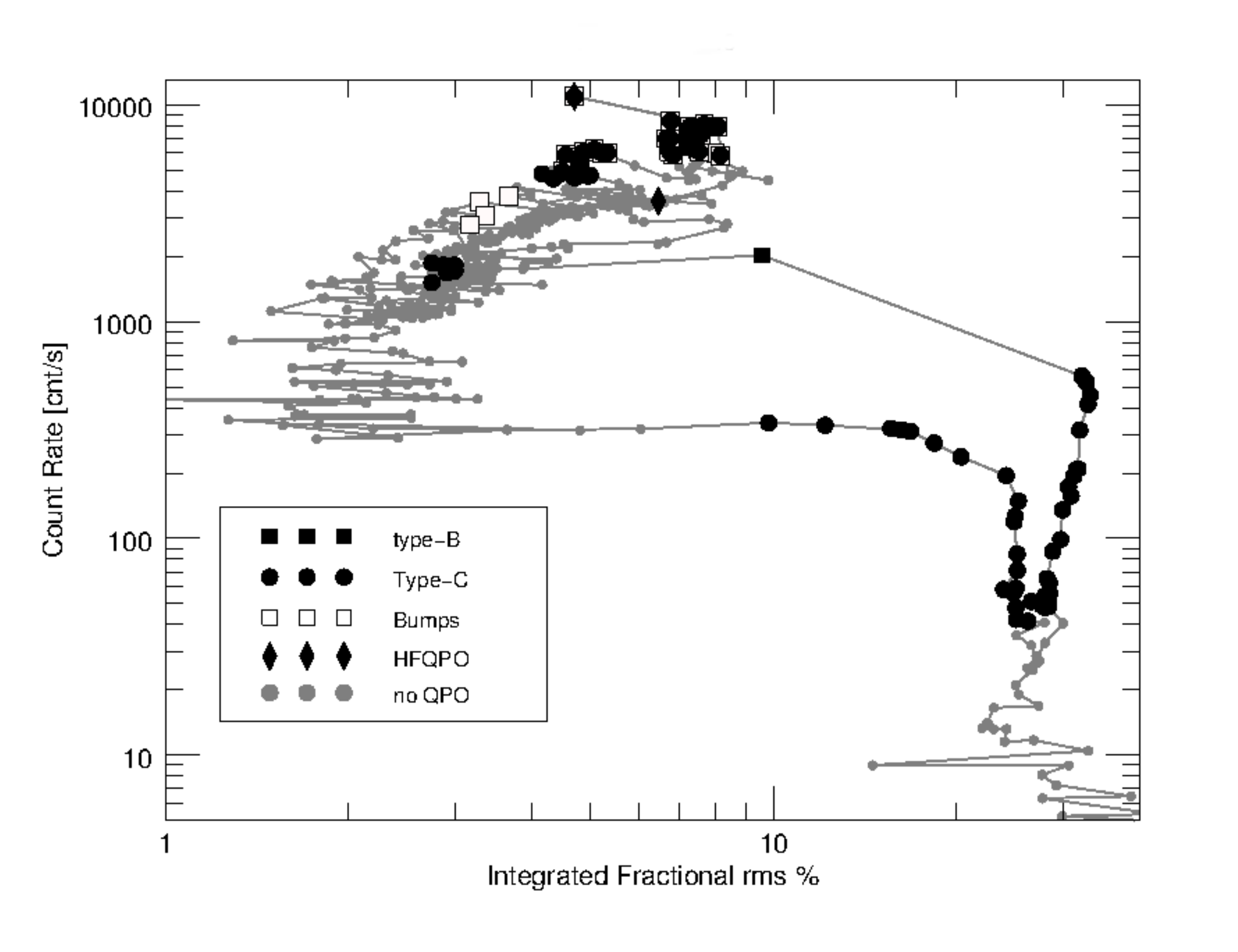}
\caption{Since the RID for BHs is less clear than the one for NSs (mainly because of a larger scatter in the intensity for different sources), we report for clarity the RID computed from GRO J1655-40 alone, where a ULS is clearly visible (outburst 2005, see \citealt{Motta2012}). Here the count rate is not rescaled.  }\label{fig:RID1655}
\end{figure}
In Fig. \ref{fig:rms_plots} (left panel) one can see that the LFQPOs from NS systems form 3 different groups:

\begin{itemize}

\item \textbf{Group 1}: this group crosses the plot from the top left to the bottom right corner, with a bend around 20-30\, Hz (circles in Fig.  \ref{fig:rms_plots}, left panel) and a sharp vertical flattening at about 50Hz. QPOs from this group cover a large frequency range ($\sim$0.1 to $\sim$60\, Hz) as well as a large rms range (3-30\%). 
QPOs from Z-sources are all found above $\sim$10\, Hz and can be all classified as HBOs, while QPOs from Atoll sources, all classified as HBO-like QPOs, cover the frequency range between 0.5 and $\sim$40\, Hz. It must be noted, however, that the largest number of HBO-like QPOs come from the Atoll 4U 1728-34, while HBO-like QPOs from other Atolls are mostly found below 10\, Hz. 

\item \textbf{Group 2}: QPOs from this group are found below Group 1, in the bottom left area of the plot and in a limited frequency ($\sim$3-9 Hz) and rms ($\sim$2-5\%) range (open and filled squares  in Fig.  \ref{fig:rms_plots}, left panel). 
This group is populated by oscillations only coming from Z-sources (included XTE J1701-462 in its Z-phase). These QPOs can be easily classified as NBOs (filled squares in the figure, see Fig. \ref{fig:QPO_sel}, right panels) or bumps (open squares).

\item \textbf{Group 3}: next to Group 2, we find Group 3, spanning the rms range 3-10\% and the frequency range 10-25\, Hz. These QPOs come from a Z source (GX 17+2, a Sco-like source) and from the Atoll with a bright Atoll phase 4U 1705-44. These QPOs can be all classified as FBOs, when coming from Z-sources, or FBO-like QPOs, when coming from the Atoll 4U 1705-44  (Fig.\ref {fig:QPO_sel}, right panels). We classified two peaks in this group (both from 4U 1705-44, open squares in Group 3, Fig. \ref{fig:rms_plots}, left panel) as bumps based on their Q (which is $<$ 2). It is worth noticing, however, that FBOs from 4U 1705-44 are generally less peaked than those detected in Z-sources and the difference between FBOs and bumps is less marked (even though broad FBOs are also sometimes seen in Z-sources, e.g. GX 349+2, \citealt{Kuulkers1998b}). Therefore, we cannot exclude that the two bumps found in group 3 are in reality low S/N FBOs (which seems likely, given their frequency). 

\end{itemize}

In Fig. \ref{fig:rms_plots} (right panel) we show the rms versus frequency plot computed for our BH sample. A comparison between the two panels in Fig. \ref{fig:rms_plots} shows that BH QPOs span narrower frequency ranges than QPOs from the NS sample, but present slightly larger scatter. The two plots evidence significant similarities between QPOs found in NS and BH systems.

\begin{itemize}

\item HBOs and HBO-like QPOs (QPOs from group 1 in the NS sample), and type-C QPOs in the BH sample are located in the same region of the rms--frequency plane, once the difference in  the frequency range covered is considered ($\sim$0.1-30\, Hz for BHs and $\sim$0.1\,Hz-60\, Hz for NS\footnote{Note that the lower limit is in general given by the frequency resolution of the data and not necessarily by physical reasons intrinsic to the source.}). These QPOs form very similar tracks in the frequency-rms plane that shows, in both cases, a clear steepening at high frequencies (around 20-30\, Hz and 10\, Hz in NSs and BHs, respectively). HBOs are located in the high-frequency half of the track, as are type-C QPOs from BHs detected either in the soft state or in the ultra-luminous state. HBO-like QPOs, instead, span the entire frequency range.
Both HBOs  and HBO-like QPOs show strong similarities with type-C QPOs for what concern their shape in the PDS (see Fig. \ref{fig:QPO_sel}). 

\item NBOs and bumps (QPOs of group 2) and type-B QPOs and bumps from BHs occupy the same region in the rms-frequency diagram for BHs, centred at $\sim$5-6Hz in both cases. 
However, in the case of NSs the rms range spanned is slightly smaller (2-5\%\, rms) than in the case of BHs (rms range 2-10\%). Also in this case, NBOs and bumps in NSs  resemble in shape type-B QPOs and bumps in BH systems, respectively (see Fig. \ref{fig:QPO_sel}).

\item FBOs and FBO-like QPOs (group 3) and type-A QPOs are also found in similar regions of the frequency-rms plane, once the difference in frequency covered is taken into account ($\sim$7\, Hz for BHs and $\sim$15\, Hz for NS). Despite slightly overlapping with type-B QPOs (thing that creates some confusion), type-A QPOs are always found at higher frequencies than type-B QPOs for a given source. This property is particularly evident when the switch between a type-B and type-A QPO occurs in matter of seconds (see, e.g., \citealt{Motta2011a} for the case of GX 339-4 and \citealt{Casella2004} for the case of XTE J1859+226), somehow similarly to the NBO to FBO transition seen in Z sources (\citealt{Dieters2000}). It musts be noted, however, that NBOs appear at slightly larger rms than type-A QPOs (3-10\%\, rms against 2-5\%).

\end{itemize}

\subsubsection{Rms-Intensity diagrams}\label{sec:RID}

In order to better comprehend the relation between the QPOs and accretion states, we built the RID following \cite{Munoz-Darias2014}, i.e. we plot the source intensity (in counts/s) versus the total integrated fractional rms for each observation and for each source in our sample, both for NSs and for BHs. 
In the case of NS systems, all count rates were rescaled to that of Cyg X-2. We used 13 kpc as a distance for Cyg X-2 (see \citealt{Munoz-Darias2014}), 9.8 kpc for GX 17+2 \citep{Galloway2008}, 3.9 kpc for Aql X-1 \citep{Galloway2008}, 8.8 kpc for XTE J1701-462 \citep{Lin2009}, 7.4 kpc for 4U 1705-44 \citep{Haberl1995}, 5.3 kpc for 4U 1728-34 (see \citealt{Galloway2008}). To date, there is no distance estimate for IGR 17473-2721, therefore, based on a similar average count rate, we assumed for this source the same distance we used for 4U 1728-34 (5.3 kpc, \citealt{Galloway2008}).
In the case of BH systems, we rescaled all the count rates at the distance of 13 kpc. We used the following distances: 6.3 kpc for XTE J1859+226 \citep{Jonker2004}, 9.1 kpc for 4U 1543-47 \citep{Orosz1998}, 8kpc for GX339-4 \citep{Hynes2004}, 4.4 kpc for XTE J1550-564 \citep{Orosz2011}, and 3.2 kpc for GRO J1655-40 \citep{Jonker2004}. We could not find a distance estimate for XTEJ1817-33 in the literature, therefore we choose 2.5 kpc in order to qualitatively reproduce the behaviour seen in the other BHBs.
We note that the uncertainty on the distance measurements would be the source of additional scatter in the points of Fig. \ref{fig:RID}. However, such scatter would be comparable to the scatter already affecting the data, therefore it would not change in the significant way the results conveyed by Fig. \ref{fig:RID}.

For simplicity, we assume here that the luminosity/count rate roughly tracks the accretion rate and we will therefore refer indifferently to one or the other depending on the context (this is, however, a simplification of a significantly more complex situation, see e.g. \citealt{Titarchuk2014}, \citealt{Seifina2015}). 
The diagrams in Fig. \ref{fig:RID} include also all those observations where no QPO are detected. These observations are marked with grey points, while the coloured points mark the different kind of QPOs, with the same colour-symbol convention used in Fig. \ref{fig:rms_plots}. 

Hysteresis cycles appear in the plot below count rates of about 100 cnts/s (corresponding to $\sim$50\% L$_{Edd}$, see \citealt{Munoz-Darias2014}). These loops cover the entire rms range, taking the sources from the hard state (rms $>$20\%), across the intermediate state (5-20\% rms), to the soft state (rms $<$5\%) and then back to the hard state through the intermediate state. This area of the diagram is formed, predictably, by the Atoll sources of our sample (4U 1705-44, Aql X-1, IGR 17473-2721,  4U 1728-34 and XTE J1701-462 in its Atoll phase). 
At higher count rates, horizontal transitions replace the hysteresis cycles, taking the source from the soft state to the intermediate state and back. This branch is populated by Z-sources (Cyg X-2 and GX 17+2), by XTE J1701-462 in its Z phase, and by the Atoll 4U 1705-44 in its bright Atoll phase, which only covers the low-variability half of this region. We will refer to areas of the plot showing hysteresis as \textit{hysteresis loops}, and to the area of the plot at high count rate with horizontal transitions and no hysteresis as \textit{high rate horizontal branch}. 

Different types of QPOs appear in different areas the RID, sometimes overlapped with one another. In particular Fig. \ref{fig:RID} (right panel) shows that:

\begin{itemize}

\item HBOs and HBO-like QPOs (circles) are found in two different regions. HBO-like QPOs are located along the hard-to-soft branch (i.e. along the top horizontal branch of the hysteresis loops) of the hysteresis loops, between 5 and 20\%\, rms, corresponding to the intermediate state. HBOs, including those from XTE J1701-462 in its Z-phase, are found on the high rate horizontal branch and cover the rms range 2-20\%, corresponding to the soft state (below 5\% rms) and to the intermediate state (5-20\% rms range). 
 
\item NBOs and bumps  (filled and empty squares, respectively) are all found along the high rate horizontal branch, mostly between 2 and 5\% rms, which corresponds to the soft state. The only point (marking a bump, open square) that is found at higher variability ($\sim$8\% rms) is from 4U 1705-44 and was located  in Fig. \ref{fig:rms_plots} (right panel), in group 3 (formed by FBOs) instead of in group 2, where all the bumps and type-B QPOs were found.

\item FBOs and FBO-like QPOs (triangles) are all found,  as NBOs, along the high rate horizontal branch between 3 and 9\%\, rms, i.e. across the transition between soft state and intermediate state. As could be seen already in Fig. \ref{fig:rms_plots} (left panel), FBOs-like QPOs (all coming from the bright Atoll 4U 1705-44) are located between 5 and 9\%\, rms, while FBOs appear at slightly lower rms (3-5\%). This is coherent with the picture outlined by \cite{Munoz-Darias2014}, where - independently from the presence of QPOs - bright Atolls only reach the low-variability end of high rate horizontal branch.

\end{itemize}

In order to compare the link between QPOs and the areas of the RID in NS and BH systems, we show in Fig. \ref{fig:RID} (right panel), the RIDs for our BHs sample. For clarity's sake, in Fig. \ref{fig:RID1655} we also report the RID from the BH LMXB GRO J1655-40, which clearly shows all the canonical spectral/timing states of BHs  (\citealt{Motta2012}), including the ULS. 
Comparing the two plots in Fig. \ref{fig:RID} we see that the RID from NS and BH LMXBs are very similar, with hysteresis loops below a certain luminosity and a high rate horizontal branch above this threshold  (\citealt{Munoz-Darias2014}). The presence of a high rate horizontal branch, corresponding to the ULS in BH systems, is however less clear in BH systems. In Fig. \ref{fig:RID} (right panel) the high rate horizontal branch is populated only by observations from GRO J1655-40 and one observation from XTE J1550-564. As already noted, however, other BH systems (not considered in this work) sometimes also cross the high rate horizontal branch.
Comparing the two RIDs we see that: 

\begin{itemize}

\item HBO-like QPOs are found on the hysteresis loops similarly to type-C QPOs from well-behaved BH systems (such as GX 339-4, that does not show an ULS, but shows clear hysteresis loops). HBOs, instead, appear only along the high rate horizontal branch, as type-C QPOs from the ULS in BH systems. However, while type-C QPOs from BHs appear all along the hysteresis loops, HBO-like QPOs only appear during the hard-to-soft branch of the hysteresis loops, and not on the soft-to-hard branch (i.e. along the bottom-half of the hysteresis loop). We note that the soft-to-hard branch takes place at significantly lower luminosities in NS systems than in BH systems, hence the detection of significant features is limited by the sensitivity of our instruments and the lack of oscillations on this branch is not surprising. Indeed, also in BHs type-C QPOs along the soft-to-hard branch show significantly worse S/N than those detected in other regions of the RID, and the detection of significant features is difficult. However, the luminosity of BHs during the soft-to-hard transition is on average still almost one order of magnitude brighter than the same transition in NSs. 

\item type-B QPOs and bumps (for both NSs and BHs), and NBOs are found in very similar rms ranges: NBOs are only detected on the high rate horizontal branch and type-B QPOs are found both on the high rate horizontal branch and on the hard-to-soft branch along the hysteresis loops.

\item type-A QPOs are located in the rms range 1-5\%, corresponding to the soft state, while FBOs are found at higher rms (5-9\%), in correspondence to the intermediate state. Additionally, while FBOs populate the high rate horizontal branch, type-A QPOs are found only on the hard-to-soft branch along the hysteresis loops. It must be noted, however, that the low number of type-A QPOs in BH LMXBs, does not allow to make a solid statement on the actual distribution of type-A QPOs in the RID.

\end{itemize}


\subsection{High frequency QPOs}\label{Sec:HFQPOs}

We now focus on kHz QPOs and HFQPOs. Following Sec. \ref{Sec:LFQPOs}, we will first analyse the rms versus frequency diagrams and then the RIDs for our NS and BH samples.
\begin{figure*}
\centering
\begin{tabular}{c c}
RMS VS FREQ DIAGRAM - NEUTRON STARS	& RMS VS FREQ DIAGRAM - BLACK HOLES \\
kHz QPOs	& HFQPOs \\
\includegraphics[width=0.49\textwidth]{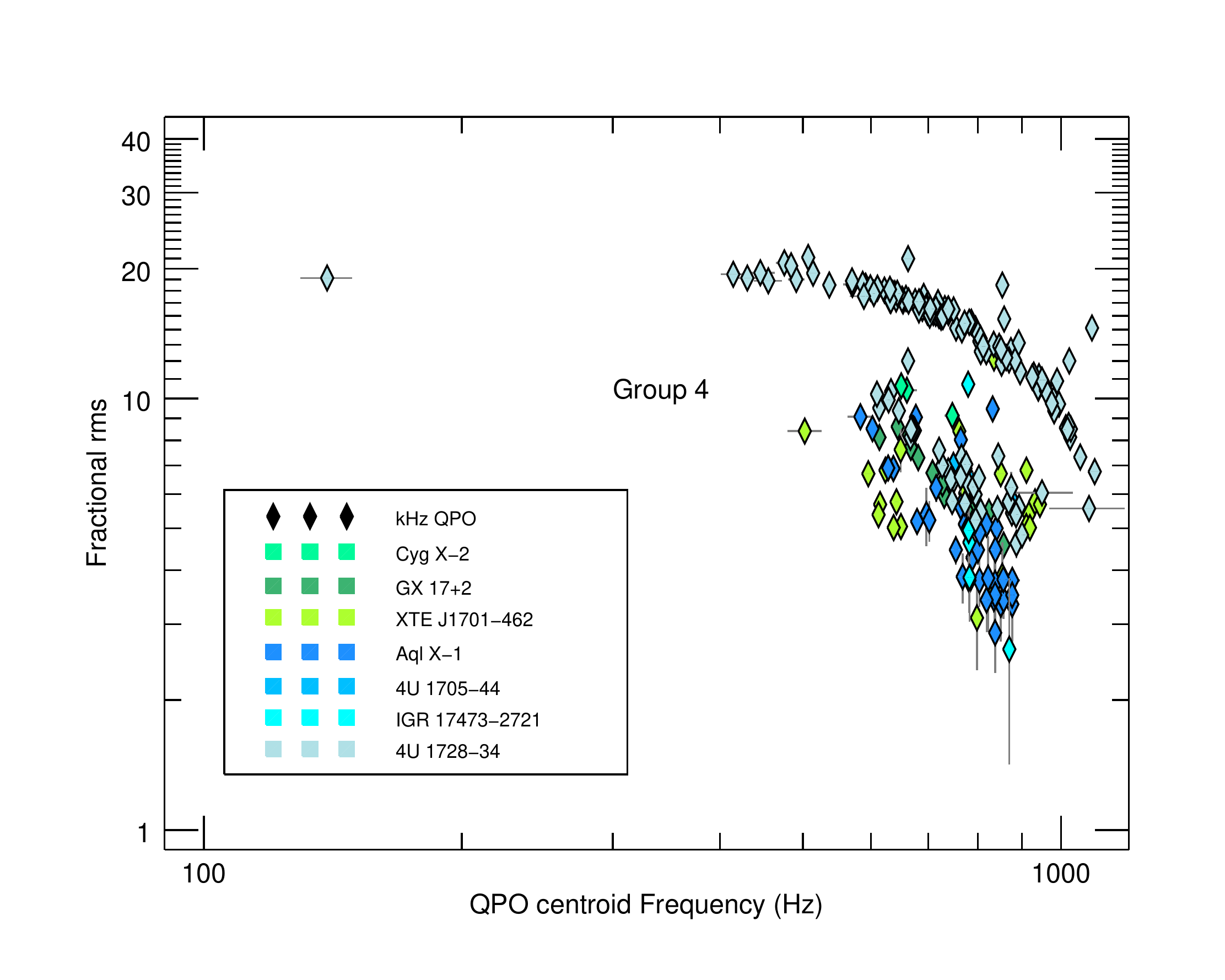} &
\includegraphics[width=0.49\textwidth]{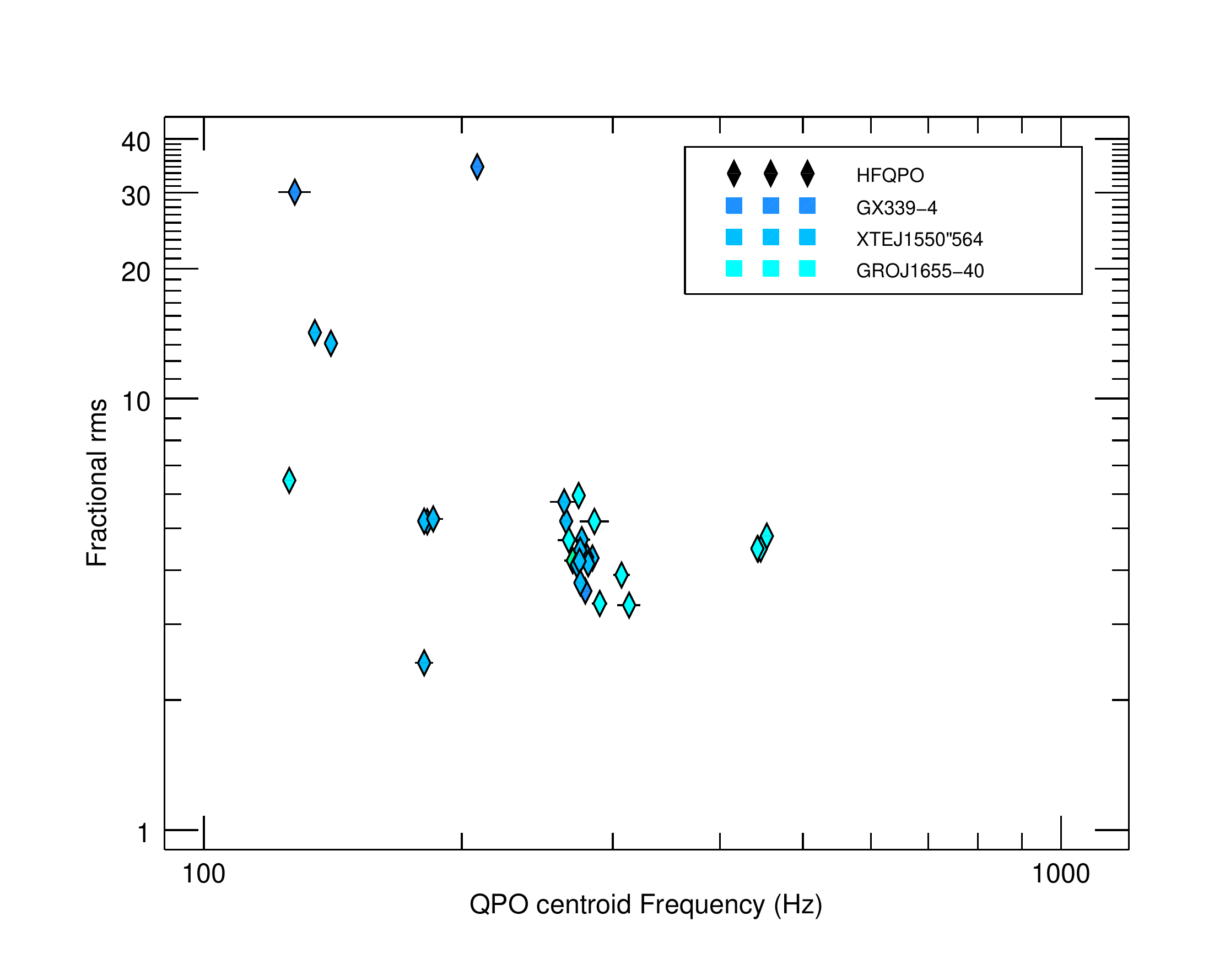} \\
\end{tabular}
\caption{\textbf{Right panel:} Integrated fractional rms versus QPO centroid frequency (as in Fig. \ref{fig:rms_plots}, left panel) for our NS sample, scaled to allow inspection of the kHz QPOs frequency region. \textbf{Left Panel:} QPO centroid frequency versus integrated fractional rms (as in Fig. \ref{fig:rms_plots}, right panel) for our BH sample, scaled to allow inspection of the HFQPOs frequency region. Frequencies are taken from \citealt{Motta2015} (LFQPOs) and \citealt{Belloni2012} (HFQPOs).}\label{fig:rms_plots_high}
\end{figure*}

%
%
%

\subsubsection{Rms-frequency diagrams}\label{sec:rms_vs_fr_high}

In Figure \ref{fig:rms_plots_high} (left panel) we show the integrated fractional rms versus the centroid frequency of all the QPOs detected in our NS sample, this time focussing on the frequency range typical of kHz QPOs.

kHz QPOs form a fourth group of points (\textbf{Group 4}) in the frequency vs rms diagram, covering the ($\sim$400, $\sim$1100) Hz frequency range, with rms lower than 20\% (diamonds in Fig.  \ref{fig:rms_plots}). This group contains both upper and lower kHz QPOs. In particular, Cyg X-2, 4U 1705-44 and Aql X-1 only show lower kHz QPOs, while GX 17+2, IGR 1747-2721, XTE J1701-462 and 4U 1728-34 show both lower and upper kHz QPOs. The continuous track at higher rms and covering the frequency range 400-1000Hz is formed only by upper kHz QPOs from 4U 1728-34.

In Figure \ref{fig:rms_plots_high} (right panel) we show the integrated fractional rms versus the centroid frequency for the HFQPOs in the BH sample. Comparing the two plots in Fig. \ref{fig:rms_plots_high}, we see that both kHz QPOs in NS LMXBs and HFQPOs in BH LMXBs populate an area of the plot at high frequency and fairly low rms. However,  while kHz QPOs form a clear correlation with rms, HFQPOs in BH XRBs form a group of point with much larger scatter, and correlation between rms and frequency is not obvious. It must be noted that we do not distinguish between upper and lower HFQPOs/kHz QPOs in the rms-frequency plots, since the classification can be problematic when only one high-frequency peak is detected. This is especially true for the case of BHs, where only GRO J1655-40 is known to show two clear simultaneous HFQPOs, which allows to identify unambiguously upper and lower peak. This fact could contribute to make the scatter of the HFQPOs population appear larger.

\subsubsection{Rms-Intensity diagrams}\label{sec:RID_high}

Fig. \ref{fig:RID} (left panel) shows that kHz QPOs (diamonds) are located in different areas of the RID. kHz QPOs from Atoll sources (Aql X-1, 4U 1728-34 and IGR 17473-2721) are located on the hard-to-soft branch of the hysteresis loops, between 5 and 10\%\,rms (corresponding to the intermediate state), and on the soft to hard transition of the hysteresis loop, covering a the rms range 3-10\% (corresponding to the soft and intermediate states). kHz QPOs from Z-sources (GX17+2 and Cyg X-2), are located, as expected, along the high rate horizontal branch, between 5 and 9\%\, rms, which corresponds to (the soft end of) the intermediate state.

A comparison with the RID build from our black hole sample (see Fig. \ref{fig:RID}, right panel) evidence that HFQPOs and kHz QPOs from Z sources are found on the high rate horizontal branch, at rms in the 2-20\% range (only two points are found at higher rms, both from GX 339-4) and 2-10\%\, rms, respectively,  therefore in the soft and intermediate states in both cases. kHz QPOs from Atoll sources, instead, are located along the hysteresis loop of the RID, where we did not detect any HFQPO.

\section{Discussion}\label{sec:discussion}

We have analysed the archival RXTE observations from 7 NS LMXBs spanning a large accretion rate range. We selected 601 observations out of 4083, where one or more QPOs and/or a peaked noise component (bump) could be detected. Our aim was to investigate the properties of quasi-periodic variability through different accretion states. 
Based on the findings by \cite{Munoz-Darias2014}, we analysed the quasi-periodic properties of NS XRBs using the fast-time variability (in the form of total fractional  rms\footnote{Differently from the QPO rms (see, e.g., \citealt{Motta2015}), the total fractional rms is essentially independent from the orbital inclination of the source, and can therefore be used to compare different sources under the spectral-states point of view.}) as a tracker of accretion state in the frequency-rms diagram and in the rms-intensity diagram (RID). We compared the QPOs found in our sample to the QPOs observed in a sample of BH LMXBs, and we found that the three types of QPOs seen in NSs show strong similarities with those detected in BHs.

\begin{figure*}
\centering
\includegraphics[width=0.98\textwidth]{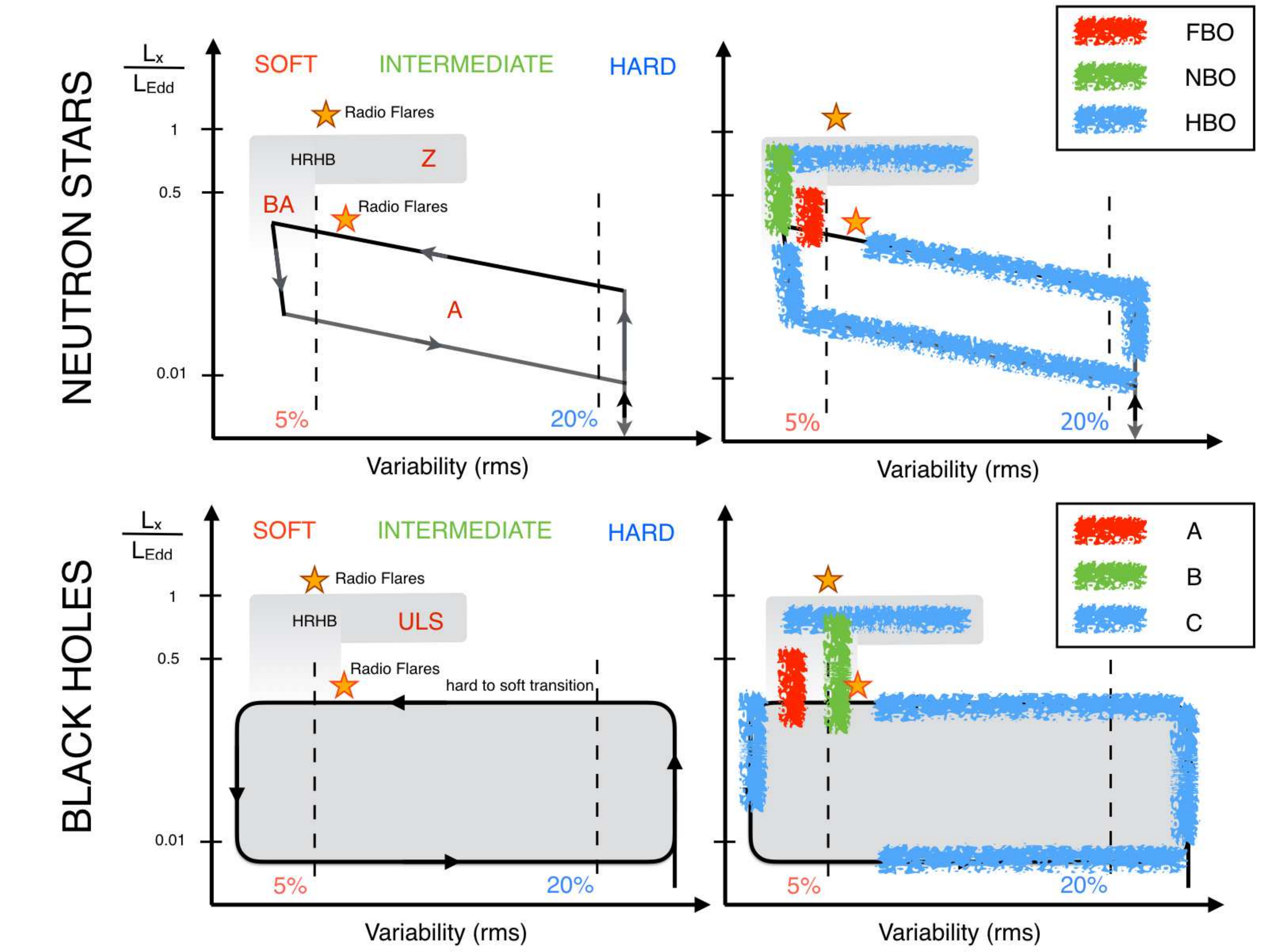}
\caption{Sketch describing the qualitative behaviour of NS (upper panels) and BH (lower panels) systems. On the left panels we show the typical behaviour for NS and BH systems, indicating the relevant spectral states. The star marks the phase where radio flares are typically observed, while the acronym HRHB stands for \textit{high rate horizontal branch.}. On the right panels, we indicate where LFQPOs are found in both systems. These sketches are to be taken as a general reference and not literally, as the behaviour of a given source may, of course, deviate slightly from the general behaviour outlined in the figures. Figures adapted from \citealt{Munoz-Darias2014}.}\label{fig:sketch}
\end{figure*}

We summarize our results in Fig.  \ref{fig:sketch}, which is based on \cite{Munoz-Darias2014} (see in particular their Fig. 9 and 10) and has been expanded including the ULS and the quasi-periodic fast-time variability properties of both NSs and BHs. Our findings broadly agree with the picture outlined by \cite{Munoz-Darias2014}, confirming it and completing it with the quasi-periodic variability information.
QPOs from both NS and BH systems follow the following pattern: HBOs and type-C QPOs are found mostly in the hard and intermediate states, while FBOs/type-A QPOs and NBOs/type-B QPOs are found close or in the soft state. HFQPOs and kHz QPOs are found in the intermediate state.

The sketches in Fig.  \ref{fig:sketch} are to be taken as a schematic view of a general picture, as the behaviour of a given source may deviate slightly from the average behaviour outlined in the figures. Specifically, while the relative luminosities marked in the figures remain valid references, the actual value of the Eddington Luminosity in different sources may significantly differ from the theoretical value, essentially dependent only on the Thomson cross section and on the mass of the compact object involved in the accretion process. \cite{Titarchuk2014} found that in Z-sources, in particular for the case of Sco X-1, the Eddington luminosity should be carefully calculated taking into account the correction to the scattering cross-section due to changes in the electron temperature kT$_{e}$ that occur when a source approaches the Eddington accretion regime. More precisely, the critical luminosity significantly increases for electron temperatures higher then kT$_{e}$ $\sim$ 50\, keV, typical of the flaring branch, during which the radiation pressure might be enough to disrupts the inner part of the accretion disk, thus causing an increase in kT$_{e}$. A similar situation is expected to occur also in bright Atoll sources (see \citealt{Seifina2015} for the case of 4U 1705-44).

\subsection{Low frequency QPOs}

\subsubsection{HBOs/HBO-like QPOs and type-C QPOs}

HBOs and HBO-like QPOs form one single track in the rms-frequency diagram, indicating that - as expected - there is no particular difference between these two QPOs populations, apart from being found in different accretion states, as it is clear from the RID.
The track formed by HBOs and HBO-like QPOs is remarkably similar to that formed by type-C QPOs in BH LMXBs, despite the larger frequency range spanned by NS systems. The fact that NSs span a larger frequency range than BHs is expected if we assume that the frequency of HBOs and type-C QPOs is somewhat dependent on the mass of the compact object, with frequencies inversely proportional to the mass (basically all models do, see, e.g., \citealt{Motta2015}). While the average BH mass is thought to be about 7M$_{\odot}$, the average NS mass is found below 2\, M$_{\odot}$, implying a factor of $\sim$3 difference in mass. Therefore, the maximum possible frequency for a BH system is expected to be lower than the maximum possible frequency for a NS system, which results in a smaller observed frequency range for BHs. We note that the minimum observed frequency for BH and NS systems is observationally the same, being imposed by our chosen frequency resolution rather than by a physical reason intrinsic to the sources.

Comparing HBOs/HBO-like QPOs and type-C QPOs, we note that they both show large frequency variations and span a wide rms range, being found in all spectral states (hard, intermediate and soft, blue areas in the plots of Fig. \ref{fig:sketch}). In  particular, both classes of QPOs can be found either at high luminosities (above 50\% L$_{Edd}$) or at low luminosities (below 50\% L$_{Edd}$, see Fig. \ref{fig:sketch}, blue areas). In particular, HBO-like QPOs and type-C QPOs from well-behaved BH systems (those showing a clear q-shaped loop in the HID or RID, such as GX 339-4) are found along the hysteresis loops, while HBOs and type-C QPOs from highly accreting BH systems (such as GRO J1665-40 in its ULS or GRS 1915+105) are detected along the high rate horizontal branch.

Based on the fact that weakly magnetized NSs and BHs are expected to have very similar effects on the surrounding space-time and consequently on matter orbiting not too close to the event horizon/NS surface, we argue that type-C QPOs in BH systems and HBOs/HBO-like QPOs in NS systems  could be the effects of the same physical process, likely sensitive to the mass of the compact object. If this is true, the physical process originating type-C QPOs and HBOs/HBO-like QPOs (regardless of what this process actually is) should be only weakly or not at all sensitive to the accretion rate, and should produce very similar effects at high or low accretion rates in order to allow both HBOs and HBO-like QPOs in NS systems as well as from high and low accretion rate BH systems to overlap perfectly in a rms vs frequency diagram, despite coming from significantly different luminosity ranges. 
An association between HBOs and type-C QPOs has been already proposed by \cite{Casella2005}, even though these authors limited their analysis to Z-sources. Our results show that the same associations can be made including HBO-like QPOs from Atoll sources. 

The relativistic precession model (RPM), proposed in original form by \cite{Stella1998} predicts that HBOs from Z-sources are the result from Lense-Thirring precession of matter around a spinning compact object. Recently, the RPM was revisited by \cite{Ingram2009, Motta2014, Ingram2013}, who showed that type-C QPOs in BH LMXBs can be explained in a similar way. According to these authors, type-C QPOs are produced via Lense-Thirring precession of the inner part of the accretion flow, whose outer radius is linked to the geometrically thin, optically thick accretion disk inner truncation radius. Thus, the size and variations of the truncation radius determine the frequency and frequency evolution the type-C QPOs (decreasing radii corresponds to increasing frequencies). In this context, the flattening occurring around 30Hz in the rms-frequency track of a few BH LMXBs (see  right panel in Fig. \ref{fig:rms_plots}, and, e.g., \citealt{Motta2012}) is interpreted as an effect of the existence of ISCO \citep{Motta2014}. The same flattening is evident in the frequency-rms track from the NS systems Cyg X-2, XTE J1701-462 and 4U 1728-34. If the Lense-Thirring nature of HBOs is correct,  the QPOs at $\sim$55\, Hz seen in Cyg X-2, for which the mass is known (M$_{NS}$ $\sim$ 1.71 M$_{\odot}$, \citealt{Casares2010}), are consistent with being produced at ISCO by a NS with adimensional spin parameter between 0.2 and 0.25, corresponds to a spin frequency of about 300-400 Hz. 

\subsubsection{NBOs and type-B QPOs}

type-B QPOs, NBOs and bumps (both from NSs and BHs) are found in the same area of the rms-frequency diagram. Coherently with the picture outlined by \cite{Munoz-Darias2014}, NBOs populate the low variability end of the high rate horizontal branch, which corresponds to the normal branch in a CCD for Z-sources (see Fig. \ref{fig:sketch}, bottom right panel, green areas).  
The fact that type-B QPOs can also be found on the hard-to-soft branch along the hysteresis loops might mean that in BHs the vertical (i.e. flux) distinction between hard-to-soft branch, the high rate horizontal branch, and the region connecting these two states (see Fig. \ref{fig:sketch}, top left panel) is less marked than in NSs. Therefore, these three regions partially overlap, allowing type-B QPOs and type-A QPOs to appear at different luminosity levels. 
This possibility is supported by the fact that while type-B QPOs are sometimes found in correspondence to peaks in the source flux (\citealt{Motta2011a} for the case of GX 339-4 and \citealt{Motta2012} for GRO J16550-40), some other times it is type-A QPOs that are observed in correspondence to flux peaks \citep{Casella2004}, similarly to what is seen for FBOs in NSs \citep{Casella2006}.

Our results confirm the association between type-B QPOs and NBO proposed by \cite{Casella2005}: based on their similarities, type-B QPOs and NBO could have similar origins. The fact that, despite the difference in mass of the compact objects involved, NBOs and type-B QPOs are found at very similar frequencies, suggests that, if type-B QPOs and NBOs are indeed produced via the same physical mechanism, either this process is only weakly sensitive to the mass of the central compact object, or different additional processes are at play, somewhat reducing/removing the dependence on the compact object mass. 

Our analysis confirms that NBO-like QPOs has never been observed so far in any Atoll source. However, our results show that bumps (observed both in Z sources and in bright Atolls) appear in the same regions as NBOs - rms, frequency and flux-wise. This might indicate a connection similar to that seen in BH systems (see \citealt{Motta2012} and \citealt{Motta2015}), where bumps become narrower and narrower as the accretion rate increases, finally evolving into a type-B QPOs. This is also true for NS Z-sources , where the bumps FWHM anti-correlates with the count rate, even though with larger scatter (\citealt{Kuulkers1998}).

\subsubsection{FBOs/FBO-like QPOs and type-A QPOs}

FBOs and FBO-like QPOs clearly form one single group in both the rms versus frequency diagram and the RID, showing no particular differences apart from the expected slightly smaller rms associated to FBO-like QPOs.  FBOs and FBO-like QPOs can, therefore, be considered part of the same populations, as already implied by \cite{Barret2002}.

Differently from the case of type-C QPOs and HBOs/HBO-like QPOs, and type-B QPOs and NBOs, which show remarkable similarities, there are a few significant differences between type-A QPOs and FBOs/FBO-like QPOs (see Fig. \ref{fig:sketch}, red areas). These QPOs are found in the same region of the rms-frequency diagram, however FBOs are centred at higher frequencies and span a larger frequency range with respect to type-A QPOs. Additionally, while type-A QPOs partly overlap with type-B QPOs in BHs in a frequency-rms diagram, there is clear separation between FBOs and NBOs in NS sources. However, as noted, when considering one source at a time, type-A QPOs always appear at higher frequency with respect to the type-B QPOs in the same source.
As already noted for the case of HBOs, the average mass of BHs is at least a factor of about 3 larger than the average mass of NSs, and the difference in frequency between type-A QPOs and FBOs is of about the same order. This might indicate that, if type-A QPOs and FBOs are produced via the same process (\citealt{Casella2005}), this process might be sensitive to the mass of the compact object (like in the case of type-C QPOs and HBOs). 

FBOs and FBO-like QPOs are also found at higher variability than type-A QPOs. This is expected, as \cite{Munoz-Darias2014} showed that NS are more variable than BHs in the soft state due to the contribution of the boundary layer to the overall emission. This is thought to be particularly relevant when the disk truncation radius approaches the ISCO. In the soft state, where type-A QPOs and FBOs are seen, the disk truncation radius is thought to be close or coincident to the ISCO. This means that if the physical process responsible for type-A QPOs and FBOs is the same, it could be enhanced (e.g. via resonance processes) or made more efficient by the presence of the boundary layer, resulting in a stronger modulation in NSs (giving rise to FBOs) than in BHs (where type-A QPOs are notoriously weak features, see e.g. \citealt{Casella2005}, \citealt{Motta2012}). 

The association between FBOs and type-A QPOs proposed by \cite{Casella2005} appears to be less obvious than the associations HBOs/HBO-like QPOs/type-C QPOs and NBOs/type-B QPOs, proposed by the same authors. Our results indicate that there are significant differences between FBOs and type-A QPOs  that should be further investigated before an association can be firmly established. However, the differences that we found are not enough to rule out such association, and based both on the similarities that we found and those reported by \cite{Casella2005}, we argue that a scenario where type-A QPOs and FBOs/FBO-like QPOs have similar origins is more likely then a scenario where there is no association between them. 

\subsubsection{Are type-A/type-B QPOs and FBOs/NBOs related?}

Our results show that, differently from HBOs, HBO-like QPOs and type-C QPOs (which appear in every state, regardless of the accretion rate), NBOs and type-B QPOs, and FBOs, FBO-like QPOs and type-A QPOs appear to require either relatively high accretion rates, a particular geometry configuration or both things at once. This is especially evident for FBOs and NBOs, that occur only  in the high rate horizontal branch , i.e. above certain luminosities (\citealt{Migliari2006}, \citealt{Munoz-Darias2014}) and in a restricted rms range. 

It is also worth noting that FBOs and NBOs are known to be linked in Z-sources. \cite{Casella2006} reported a smooth (fast but resolved) transition from a FBO to an NBO in the RXTE data of Sco X-1 (see also \citealt{Priedhorsky1986}, \citealt{Dieters2000}, \citealt{Titarchuk2014}). 
This transition is very different from the fast, unresolved transitions observed between type-B and type-A QPOs in BHs (\citealt{Nespoli2003}, \citealt{Casella2004})\footnote{Note that the transition in both NSs and BHs can go either way, i.e. type-B to type-A or \textit{vice versa} and FBO to NBO or \textit{vice versa}.}. However, the difference in the two transitions could be just a time-scale one, with the transition in BHs being identical to that in NSs, but occurring on a much shorter time scale. This would imply a link between type-A and -B QPOs similar to the link between FBOs and NBOs. Should this link be confirmed, it would provide further support to the the association type-A QPOs/FBOs discussed above. 
Future missions, with larger effective areas than RXTE (such as eXTP, http://isdc.unige.ch/extp/), will be able to resolve the type-A/B transition in BH LMXBs, allowing to confirm or reject the similarity with the FBO/NBO transition in NS binaries.

\subsection{High frequency QPOs}

HFQPOs in BHs and kHz QPOs in NSs populate the same region of the rms-frequency diagram, with NS system reaching higher frequencies. 
While in NSs the kHz QPOs form a clear frequency-rms track, this is not true for BHs, where we see significantly larger scatter. Independently on the process originating kHz QPOs and HFQPOs, their frequency are likely related to the mass of the compact object of their host system (see, e.g., \citealt{Torok2012}). Therefore, this scatter might be explained as a consequence of the wider mass distribution of BH in LMXBs ($\sim$3-20 M$_{\odot}$) with respect to NSs ($\sim$1.4-2.8 M$_{\odot}$). In the same way, the larger average BH mass in BH systems compared to the average mass of NS probably explains the fact that HFQPOs reach lower frequencies than kHz QPOs.  

The most significant difference between HFQPOs in BHs and kHz QPOs in NSs is that while in BHs HFQPOs only appear at the highest accretion rates, i.e. along the high rate horizontal branch (see also \citealt{Motta2016} and \citealt{Belloni2016}), in NS systems kHz QPOs are located both along the hysteresis loops and on the horizontal high-count rate branch. This indicate that, contrary to what has been speculated for BH systems (see e.g. \citealt{Belloni2016}), an high accretion rate might not be a necessary condition for the appearance of kHz QPOs in NS. We must, however, consider the fact that the lack of HFQPOs in BH systems at low luminosities could be a selection effect due to the intrinsic smaller amplitude of HFQPOs in BHs with respect to that of kHz QPOs in NSs. 
 
In both NS and BH systems, HFQPOs and kHz QPOs are consistently observed in the intermediate and soft state (with only two exceptions in the case of BHs, where two HFQPOs, both from GX 339-4, are observed at $\sim$30\% rms). 
Therefore, if HFQPOs and kHz QPOs are the result of the same process (as it has been speculated since the discovery of HFQPOs in BH LMXBs, see \citealt{Morgan1997} and \citealt{Strohmayer2001}), then such process should not be (too) sensitive to the mass accretion rate, but could be instead triggered when a certain geometry is reached and/or certain physical conditions in the accretion flow are met (see e.g. \citealt{Zhang2017}).

\subsection{Caveats}

The NS sources we considered in this work constitute only a sample of all the NS systems observed by RXTE \citep{Munoz-Darias2014}. Even though our sample can be reasonably considered representative of the behaviour of a much larger population of NS LMXBs,  a study such as that presented here on an independent sample of sources (ideally on all the remaining NS LMXBs) would be desirable to further confirm our findings. However,  we stress that our sample contains a number of Atolls and Z sources that behave consistently within their class under a quasi-periodic variability point of view, while were already proven to behave consistently from the accretion states point of view in a picture valid for BH systems  as well \citep{Munoz-Darias2014}. Thus, there is no particular reason to believe that our results would not be consistent with those obtained from a larger number of sources.

\section{Summary and conclusions}

We reported on the systematic study of QPOs detected in a sample of neutron star X-ray binary systems, including Atoll and Z-sources and spanning a large accretion rate range. We treated in the same way all the sources of our sample, without assuming \textit{a priori}  that any particular property of either class (Atoll or Z) would influence the quasi-periodic variability of the systems. 
We could confirm the association of NBOs and HBOs/HBO-like QPOs to type-B and type-C QPOs, respectively, already proposed in the past but limited to Z-sources and BH systems. While the similarities between HBOs/HBO-like QPOs and type-C QPOs, and NBOs and type-B QPOs are remarkable, the similarities between type-A QPOs and FBOs/FBO-like QPOs are less obvious, although our results still support the association of these two types of QPOs. %
Based on our findings, it would be in principle possible to apply the same classification to the LF and maybe HF QPOs from both NS and BH LMXBs, in order to match the common, variability-based, state classification scheme that we used in this work.

We added to the picture outlined by \cite{Munoz-Darias2014} the quasi-periodic variability properties of both BH and NS systems. Our analysis shows that in both NS and BH systems, different accretion rates do not seem to affect the main properties of HBOs/type-C QPOs and kHz QPOs/HFQPOs, which appear to behave in the same way during low accretion states (during the hysteresis loops) and at significantly higher accretion rates (along the high accretion rate horizontal branch). On the contrary, we have shown that FBOs/NBOs and type-A/B QPOs require fairly high accretion rates or particular geometry configurations, or both,  to be seen.

Finally, our results confirm that BH and weekly-magnetized NS binary systems are remarkably similar not only in their hysteresis/state changes behaviour, but also in their quasi-periodic variability properties, which appears to be only weekly sensitive to the nature of the compact object involved in the accretion process.
\bigskip
\bigskip

\small
\noindent \textit{SEM thanks Jeroen Homan and Marike van Doesburgh for useful discussion, and the anonimous referee whose comments and suggestions contributed to improve the quality of this work. SEM and ARE acknowledge the support from the ESAC Faculty. SEM acknowledges the University of Oxford and the Violette and Samuel Glasstone Research Fellowship program. TMD acknowledges support via a Ram\'on y Cajal Fellowship (RYC-2015-18148).
This research has made use of data obtained from the High Energy Astrophysics Science Archive Research Center (HEASARC), provided by NASA's Goddard Space Flight Center.}
\normalsize

\newpage

\appendix
\section{The case of 4U 1728-34}\label{sec:4U1728}

While most the sources of our sample show fairly clear PDS, where LFQPOs appear as one, easily identifiable strong peak (the fundamental, or first harmonic) sometimes accompanied by significantly weaker overtones (or second, fourth and so on harmonics), this is not the case for 4U 1728-34. The PDS from this source often show a number of narrow peaks in a harmonic relation, and it is not always clear which one is the fundamental.

\begin{figure}
\centering
\includegraphics[width=0.5\textwidth]{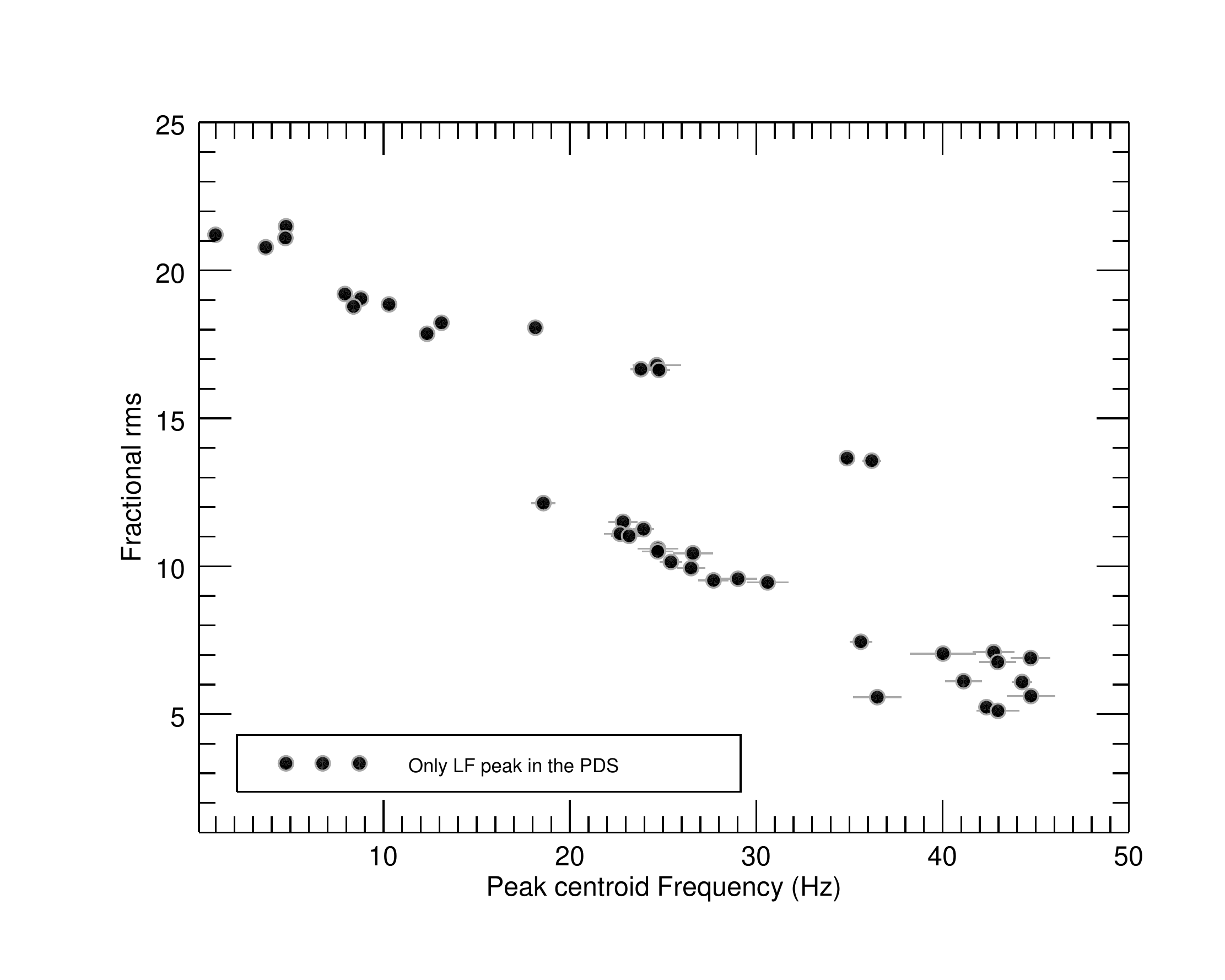}
\includegraphics[width=0.5\textwidth]{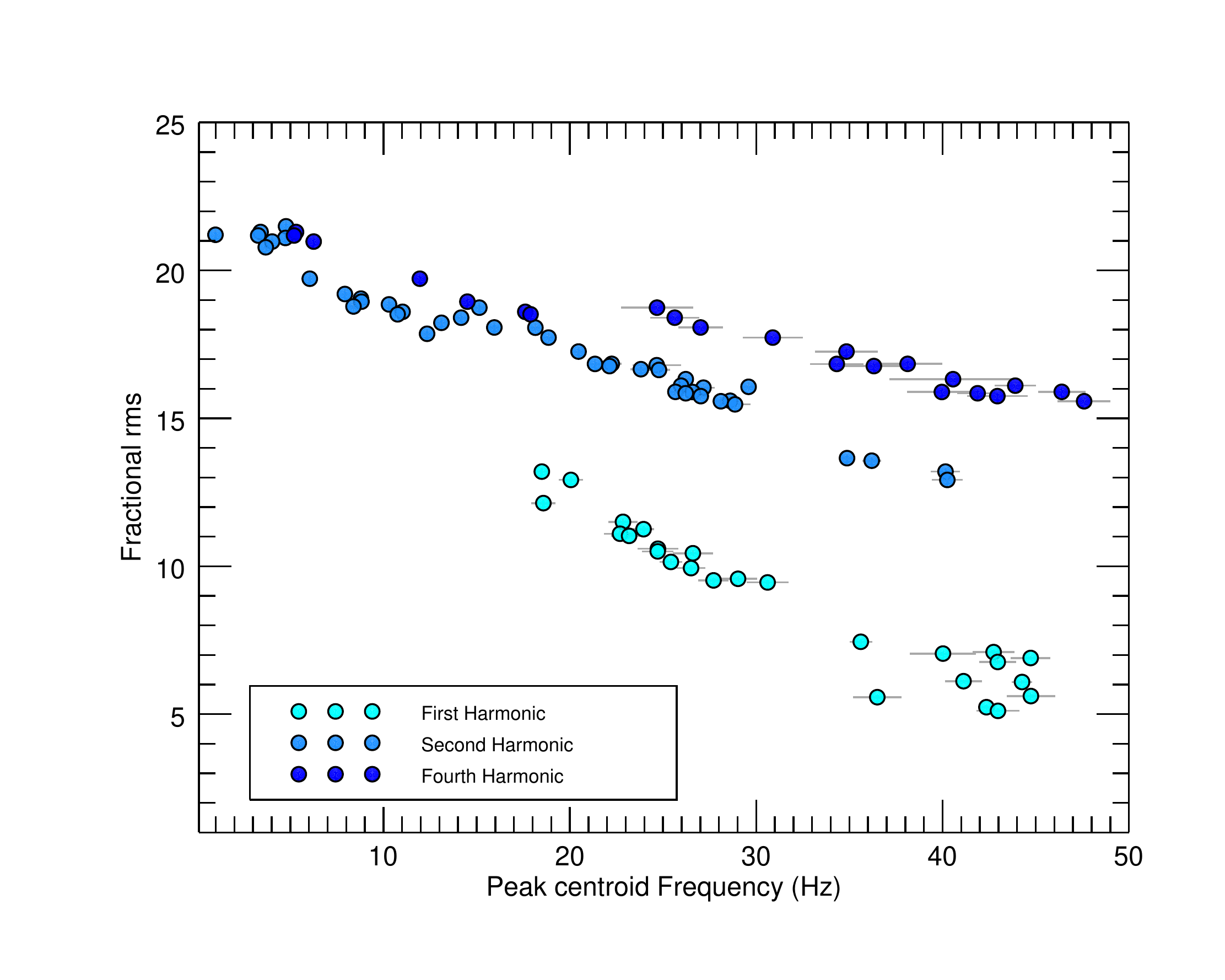}
\includegraphics[width=0.5\textwidth]{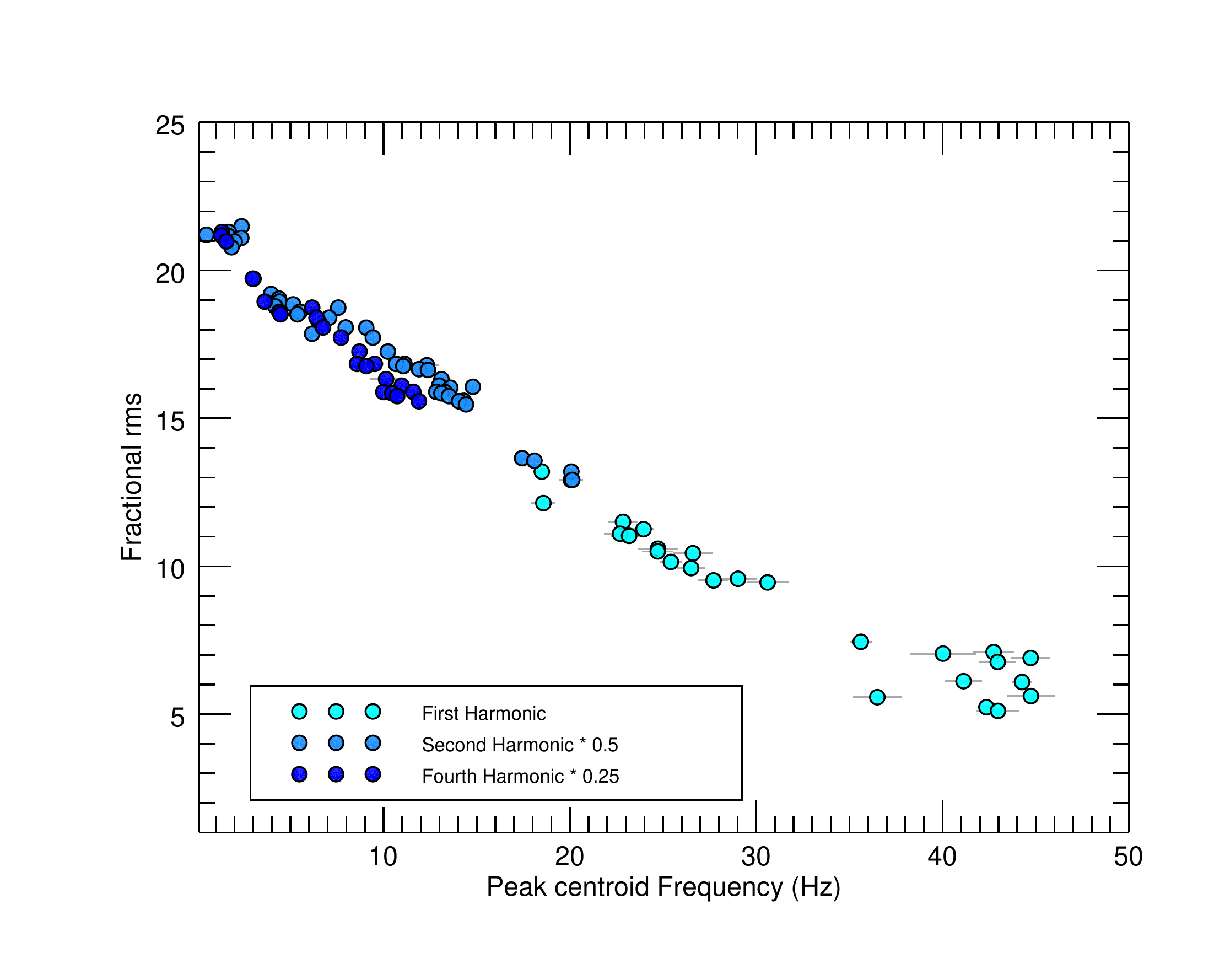}

\caption{Integrated fractional rms plotted against the frequency of each low-frequency QPO peak in the PDS from 4U 1728-34.}\label{fig: 1728}
\end{figure}

As we aim at the fundamental frequency only, in order to pick the correct QPO peak, we plotted the rms\footnote{The integrated fractional rms is known to tightly correlate with the characteristic frequencies of the PDS, i.e. break frequency, HBO frequency, kHz QPOs frequency, see e.g. \citealt{Casella2005}} as a function of all the low-frequency peaks from each PDS (see Fig. \ref{fig: 1728}). We started by plotting only those QPOs that show only one peak (i.e. no harmonics are significantly detected, see Fig. \ref{fig: 1728}, top panel). We see that the rms versus frequency tracks divides in two branches: one spanning the $\sim$1-36\, Hz frequency range, above $\sim$13\%\, rms; another spanning the $\sim$18-45\, Hz frequency range, below 13\%\,rms. These branches seem to be in harmonic relation (extrapolating at higher and lower frequencies) and we judge that the branch below 13\% is formed by first harmonics, while the branch above 13\% rms is formed by second harmonics. Even if we cannot exclude that the  branch above 13\% rms is formed by first harmonics (hence the branch below 13\% rms would be formed by sub-harmonics), we consider this unlikely since this would imply a maximum fundamental frequency of $\sim$90 Hz, higher than any HBO frequency found in the literature. 

Following the same approach, we then added to the plot all the QPOs coming from cases where two peaks are seen\footnote{In a few cases also a third peak is visible, but it is never formally a QPO peak, since its Q is lower than 2, therefore we did not consider those peaks here.} (Fig. \ref{fig: 1728}, middle panel). 
As in the previous case, the lower frequency peaks divide in two tracks, one above and one below 13\%\,rms, which - based on what discussed above - are formed by second harmonics and first harmonics, respectively. This means that the higher frequency peak of each couple is the second harmonic from those QPOs found below 13\%\,rms, or a fourth harmonic for those QPOs detected above 13\%\,rms. We note, however, that while the harmonic relation between first and second harmonic is neat, the fourth harmonic does not always clearly respect the harmonic series, deviating slightly from the expected frequencies (see \citealt{vanDoesburgh2017}). Since we are only interested in the behaviour of the fundamental frequency, this fact does not affect our analysis and further investigations are beyond the scope of this work. We will extensively study the properties of the harmonic content in BH and NS systems in a forthcoming work (Motta \& Ingram in prep).

Based on what discussed above, in order to compare the QPOs from 4U 1728-34 with those from other systems, we will refer only to the fundamental frequency. Therefore, when the fundamental frequency is not detected in the PDS, we report the frequency inferred from the second harmonic (or first overtone), based on the classification outlined above. The frequencies (measured or inferred) used in Fig. \ref{fig:rms_plots} (left panel) and \ref{fig:RID} (left panel) are plotted as a function of rms, are shown in Fig. \ref{fig: 1728} (bottom panel). 

\bigskip

\bibliographystyle{mn2e.bst}
\bibliography{biblio}

\label{lastpage}
\end{document}